\documentclass[twocolumn]{aastex63}
\usepackage{xcolor}
\usepackage{makecell}
\usepackage{pbox}
\usepackage{amsmath}

\begin{document}

\title{Population Synthesis of Black Hole Binaries with Compact Star Companions}

\author{Yong Shao}
\affiliation{Department of Astronomy, Nanjing University, Nanjing 210023, People's Republic of China; shaoyong@nju.edu.cn}
\affiliation{Key laboratory of Modern Astronomy and Astrophysics (Nanjing University), Ministry of
Education, Nanjing 210023, People's Republic of China; lixd@nju.edu.cn}

\author{ Xiang-Dong Li}
\affiliation{Department of Astronomy, Nanjing University, Nanjing 210023, People's Republic of China; shaoyong@nju.edu.cn}
\affiliation{Key laboratory of Modern Astronomy and Astrophysics (Nanjing University), Ministry of
Education, Nanjing 210023, People's Republic of China; lixd@nju.edu.cn}

\begin{abstract}

We perform a systematic study of merging black hole (BH) binaries with compact star (CS) companions, including
black hole--white dwarf (BH--WD), black hole--neutron star (BH--NS) and black hole--black hole (BH--BH) systems. 
Previous studies have shown that 
mass transfer stability and common envelope evolution can significantly affect the formation of merging BH--CS binaries 
through isolated binary evolution. With detailed binary evolution simulations, we obtain easy-to-use criteria 
for the occurrence of the common envelope phase in mass-transferring BH binaries with a nondegenerate donor, and 
incorporate into population synthesis calculations. To explore the impact of possible mass gap between NSs and BHs on the 
properties of merging BH--CS binary population, we adopt different supernova mechanisms involving the \textit{rapid},
\textit{delayed} and \textit{stochastic} prescriptions to deal with the compact remnant masses and the natal kicks. 
Our calculations show that there are $ \sim 10^{5} -10^{6}$
BH--CS binaries in the Milky Way,  among which dozens are observable by future space-based gravitational wave detectors. 
We estimate that the local merger rate density of all BH--CS systems is $ \sim 60-200 \,\rm Gpc^{-3}yr^{-1}$.  
While there are no low-mass BHs formed via \textit{rapid} supernovae,
both \textit{delayed} and \textit{stochastic} prescriptions predict that $ \sim 100\% $/$ \sim 70\% $/$ \sim 30\% $ of 
merging BH--WD/BH--NS/BH--BH binaries are likely to have BH components within the mass gap.

\end{abstract}

\keywords{Gravitational waves -- Compact binary stars -- Black holes -- Stellar evolution -- Supernovae}

\section{Introduction}

Stars are believed to end their lives as compact stars (CSs) that include white dwarfs (WDs), neutron stars (NSs) and
black holes (BHs). Observations indicate that massive stars are predominately born in binary and multiple systems 
\citep{sd12,kk14,md17}. It is naturally expected that CS pairs can be formed through isolated binary evolution when both
components have evolved to be CSs.
Very close CS pairs emit gravitational wave (GW) signals that are likely to be identified by future space-based detectors e.g. LISA 
\citep{asa17}, TianQin \citep{lj16} and Taiji \citep{rw20}, and/or by ground-based interferometers e.g. LIGO \citep{aa15} and 
Virgo \citep{ac15}. Common envelope \citep[CE, see a review by][]{in13} evolution is thought to be a vital stage for the formation
of close CS pairs, during which the binary systems rapidly shrink inside a shared envelope originating from the donor star
and the orbital energy is dissipated to expel the CE \citep{p76,w84,il93}. Usually dynamically unstable mass transfer between 
binary components can result in the occurrence of a CE phase, which is critically dependent on the stellar properties
of both the donor and the accretor, and the mass ratios of binary components \citep[e.g.,][]{sp97,ge10,ge20,sl14,pi15,pi17}.
It is generally accepted that a binary system is likely to enter CE evolution if the mass ratio of binary components
is extreme or the donor star has developed a deep convective envelope prior to the mass transfer. 

Radio and X-ray observations of Galactic NSs suggest a maximum mass of $ \sim 2M_{\odot} $ \citep{af13,asb18,fc20}, 
which is consistent with the
maximum stable NS mass inferred from observations of the  NS--NS merger GW170817 \citep{mm17,rs18,sz19}. 
Meanwhile, Galactic BHs in X-ray binaries are inferred to have a minimal mass of $ \sim 5M_{\odot} $ \citep{bj98,op10,fs11}. This
led to the mass gap ($ \sim 2-5M_{\odot} $) between NSs and BHs. However, recent evidence suggest that the mass gap 
is being populated from both electromagnetic \citep{tk19,wm20,js21} and GW observations 
\citep[GW190814,][]{ab20a}. Whether there is the mass gap or not can shed light on supernova 
mechanisms for the formation of NSs and BHs. Although Gaia astrometry was already claimed to be able to
discover invisible BHs with optical companions \citep[e.g.,][]{gs02,ml17,bcl17,yb18,yk18,bca19,sl19,abc19,wl20},
merging BH--CS binaries that appear as   
GW sources are alternatively potential objects to test the mass gap and relevant supernova mechanisms. 

There is no BH--WD binary identified so far, although they are proposed to be associated with many kinds of astrophysical 
objects, e.g. ultracompact X-ray binaries \citep[UCXBs,][]{vh12},  subluminous type I supernovae \citep{mbd12}, 
tidal disruption events \citep{fm20}, and long gamma-ray bursts \citep{dg18}. Some X-ray binaries in globular
clusters are suggested to be BH--WD systems \citep{mk07,mj15}, and the formation of such systems 
in a dynamical environment has been explored by \citet{ic10}. It is also noted that mass-transferring BH--WD binaries
are potential GW sources that may be identified by LISA in the future \citep{st21}. 

Since the discovery of the first GW signal from a BH--BH inspiral \citep{ab16}, the LIGO/Virgo collaboration 
has reported the detection of tens of  CS pair mergers \citep{ab19,ab20b}. Among them the majority are confirmed to be BH--BH 
systems and several are suspected to be BH--NS systems, e.g. GW190814 \citep{hh20,zl20} and GW190425 \citep{ht20,kf20}. 
As expected,  
searches of the electromagnetic counterparts for BH--BH mergers have yielded negative results \citep[e.g.,][]{ct16,sc16,rb17}. 
It is predicted that LISA will detect a number of merging BH--BH binaries in the Milky Way, but that electromagnetic
observations will be challenging \citep{lg18}.

The detection of BH--NS mergers is of great interest as they are expected to emit across a broad electromagnetic spectrum 
and have been suggested to produce radio flares \citep{np11,hn16}, kilonovae \citep{lp98,zw20}, short 
gamma-ray bursts \citep{p86,gl20}, and so on. A recent quite comprehensive study of BH--NS mergers can be found in \citet{bbn21}.
Before merger, Galactic BH--NS binaries are possibly observed as binary 
radio pulsars \citep{sl18}.  Close BH--NS systems in the Milky Way may emit GW signals
with frequencies in the LISA band \citep{cs20}. 

With the growing population of GW sources, a large number of formation channels for merging CS pairs
have been put forward, especially for the case of BH--BH systems. Until now, it is still impossible to distinguish between various 
channels based on GW observations alone. The popular channels involve the formation 
through isolated binary evolution 
\citep[e.g.,][]{ty93,lp97,vt03,ki14,bh16,es16,sv17,vdh17,km18,mg18,sm19,bc20,zs20,ts21,bf21,ob21}, and dynamical 
interactions in globular clusters \citep{db10,rh16,as17,pw19,ky20} or young stellar clusters 
\citep{zm14,dgm19,sm20,rm20}. Other channels include the formation from isolated multiple systems 
\citep{st17,ll18,hn18,fl19}, the chemically homogeneous 
evolution for rapidly rotating stars \citep{md16,dm16,ml16}, as well as the disk of active galactic nuclei \citep{ar16,sm17,mf18}. 

In this paper, we investigate the formation of merging BH--CS binaries via isolated binary evolution. Our previous works
have focused on the Galactic populations of detached BH binaries with a normal-star companion \citep{sl19} 
and BH X-ray binaries \citep{sl20}, by adopting  the rapid \citep{fb12} and the failed supernova mechanisms  
\citep{se16,rso18} for NS/BH formation. Here we further incorporate the delayed \citep{fb12} and the stochastic \citep{mm20} recipes
to deal with compact remnant masses and possible natal kicks, and then study the predicted properties of 
descendent BH binaries with a CS companion. 
For mass-transferring binaries with a BH accretor, \citet{pi17} demonstrated that the process of Roche lobe overflow (RLO) may be 
stable over a much wider parameter space than previously thought. Accordingly, \citet{ob21} suggested that sufficiently strong 
constraints on mass transfer stability are necessary to draw fully reliable conclusions for the population of double CS mergers. 
However, considering that limited binary systems were included by \citet{pi17}, 
in this work we will evolve a large grid of the initial parameters for BH binaries
with a nondegenerate donor to deal with mass transfer stability and
obtain thorough criteria (or parameter spaces) for the occurrence of CE evolution. 

The structure of this paper is organized 
as follows.  Section 2 describes the method of our binary population synthesis \citep[BPS, see a review by][]{hg20}
and the input physics implemented
in the BPS method, especially supernova mechanisms for NS and BH formation. In Section 3 we obtain the criteria 
of mass transfer stability for the BH binaries with detailed binary evolution calculation. Sections 4 and 5 show our BPS 
calculation results about the merging BH--CS binaries in the Milky Way and the local Universe, respectively. 
Finally we conclude in Section 6.

\section{BPS Method}

To simulate the formation and evolution of BH--CS binaries, we utilize the \textit{BSE} code originally developed by 
\citet{h02} and significantly updated by \citet{sl14}. Some further modifications on the code can be found in \citet{sl19}
and \citet{sl20}. We briefly summarize the most important points in the following. When dealing with the process of mass transfer  
in the primordial binaries, we use the rotation-dependent mode \citep{sl14} that assumes the accretion rate 
of the secondary stars to be dependent on their rotational velocities \citep[see also][]{se09}. With this mode we are able to reproduce
the distributions of known Galactic binaries including BH--Be star systems \citep{sl14,sl20} and 
Wolf-Rayet star--O-type star systems \citep{sl16}.  Because a large fraction of the transferred mass is lost from the rapidly rotating
secondary star, the maximal mass ratio of the primary to the secondary stars for stable mass transfer can reach up to $ \sim 6 $, 
significantly larger than in the conservative mass transfer case \citep{sl14}. 
During CE evolution, we employ the
binding energy parameter $ \lambda $ calculated by \citet{xl10} and set the CE ejection efficiency $ \alpha_{\rm CE} $ to be unity.
We follow \citet{bb10} to deal with the wind mass-loss rates for different types of stars, except that for helium stars we decrease 
the mass-loss rates of \citet{ham95} by a factor of 2 \citep{kh06}. 

For the formation of NSs and BHs, we take into account three supernova models to treat the remnant masses 
and natal kicks. These models involve (i) the rapid explosion mechanism \citep{fb12}, (ii) the delayed explosion mechanism \citep{fb12} 
and (iii) the stochastic recipe developed by \citet{mm20}. The rapid model predicts a dearth in the remnant
masses between $ \sim 2 M_{\odot} $ and $ \sim 5 M_{\odot} $, while the other two models are able to produce CSs within the mass gap.
For both the rapid and the delayed mechanisms, the remnant masses are determined by the CO core masses at the time of explosions,
and subsequent accretion of the fallback material. 
We follow \citet{fb12} to convert between the baryonic and gravitational masses for NSs \citep[see also][]{tww96}
and simply approximate the gravitational mass with $ 90\% $ of the baryonic mass for BHs. 
The maximum mass of NSs is set to be $ 2.5M_{\odot} $. We adopt the criterion
of \citet{fb12} to distinguish the NSs originating from core-collapse and electron-capture supernovae. 
NSs from core-collapse supernovae are assumed to be subject to a kick with a Maxwellian distribution with
$ \sigma = 265 \rm \,km\,s^{-1} $ \citep{h05}, while NSs from electron-capture supernovae
have a lower kick velocity with $ \sigma = 50 \rm \,km\,s^{-1} $ \citep{db06}. For the natal kicks to newborn BHs,
we use the NS kick velocities reduced by a factor of $ (1-f_{\rm fb} )$, where $ f_{\rm fb} $ is the fraction of the
fallback material. In the stochastic model, the outcome of supernova explosions is expected to be
probabilistic rather than deterministic. The remnant masses and natal kicks are required to satisfy some specific 
probability distributions depending on the masses of the CO cores. Meanwhile, the hydrogen shell (if present) 
is assumed to be always ejected, and this will cap the BH masses at the helium core masses. 
This model allows a significant tail of low kicks
for natal NSs, which can be consistent with the observation of large numbers of  NSs in globular clusters \citep{mm20}.
In addition, a large fraction of BHs are expected to receive zero natal kicks. 
Following \citet{mm20}, we take the 
maximum mass of NSs to be $ 2M_{\odot} $\footnote{The maximum mass of observed NSs is a bit higher, around $ 2.1M_{\odot} $, but
this does not influence our final results.}.

\begin{figure}[hbtp]
\centering
\includegraphics[width=0.45\textwidth]{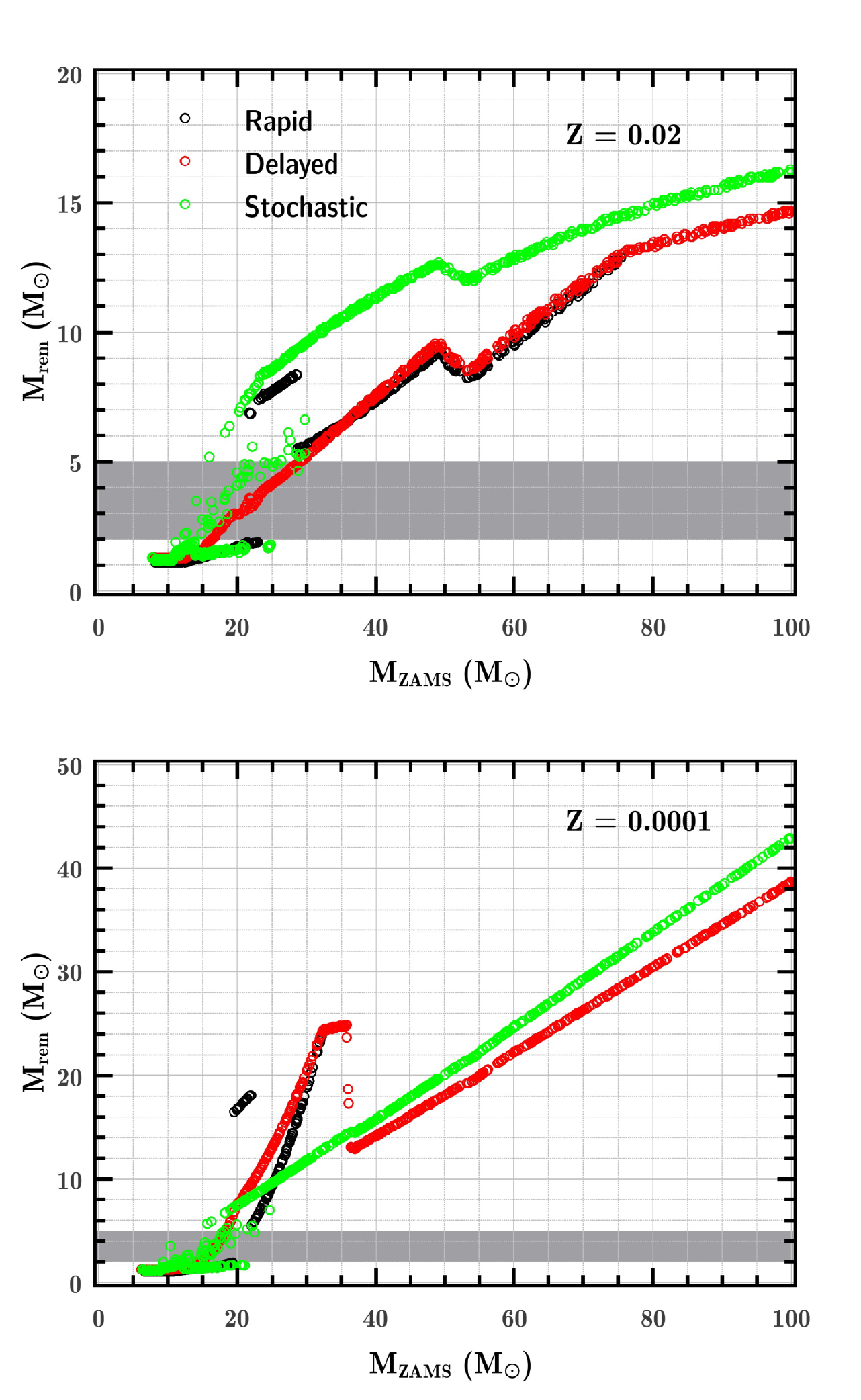}
%\linespread{0.7}
\caption{The remnant mass as a function of the zero-age main sequence mass of the progenitor stars for our adopted three supernova mechanisms. 
These results are obtained by using the population synthesis calculations for single stars. The top and bottom panels correspond to the  stars
with initial metallicity of  0.02 and 0.0001, respectively. 
The gray rectangle in each panel highlights the mass gap. 
   \label{figure12W}}
\end{figure}

Figure 1 shows the relations between the zero-age main sequence mass and the remnant mass for our adopted 
three supernova models. For the models involving the rapid and delayed explosion mechanisms, similar 
relations have been reported by many previous works \citep[e.g.,][]{bb10,fb12,gm18,sl19,zs20}. 
It is noted that the minimum
mass of the BH's progenitors can reach as low as $ \sim 12M_{\odot} $ in the stochastic model. Since we 
follow \citet{mm20} to ignore the correction between the baryonic and gravitational masses, 
the stochastic model tends to produce slightly heavier BHs at the high mass end compared with those in the other two models.
In our BPS calculations, we identify mass-gap BHs with masses between $2M_{\odot}$  and $5M_{\odot} $ for all models.

The evolutionary fate of the primordial binaries is determined by their initial parameters: the primary masses $ M_{\rm 1} $,
the secondary masses $ M_{\rm 2} $, the orbital separations $ a $ and the eccentricities. Since the eccentricity
distribution has a minor effect on the BPS results \citep{h02}, we assume that all primordial binaries have circular orbits for simplicity.
In our calculations, we take  $ M_{\rm 1} $  in the range of $ 5-100 M_\odot $,  $ M_{\rm 2} $ 
in the range of $ 1-100 M_\odot $, and  $ a $ in the range of $ 3-10^4 R_\odot $. 
For the secondary stars, only the binaries with $ M_{2} < M_{1} $ are included. 
We set each of the initial parameters $ M_{1} $, $ M_{2} $ and $ a $ with the $ n_{\chi} $ grid points of the
parameter $ \chi $ logarithmically spaced, so
\begin{eqnarray}
\delta \ln \chi = \frac{1}{n_{\chi}-1} (\ln \chi_{\rm max} - \ln \chi_{\rm min}).
\end{eqnarray}
Here $ n_{\chi} $ is taken to be 100.
If a specific binary $ i $ evolves across a phase that is identified as a BH--CS system, then the binary  
contributes the BH--CS binary population with a rate
\begin{eqnarray}
R_{i} =  \left( \frac{f_{\rm bin}}{2} \right) \left( \frac{\rm SFR}{M_{*}} \right) W_{\rm b},
\end{eqnarray}
where  $ f_{\rm bin} $ is the binary fraction,  $ \rm SFR $ is the star formation rate,  
$M_{*} \sim 0.5 M_{\odot}$ is the average mass for all stars, and 
$W_{\rm b} = \Phi (\ln M_{1}) \varphi (\ln M_{2})
\Psi (\ln a) \delta \ln M_{1} \delta \ln  M_{2} \delta \ln  a$
weights the contribution of the specific binary from the primordial binary with 
initial parameters of $ M_{1} $, $  M_{2} $ and $ a $ \citep[see more details in][]{h02}.
We assume that all stars are initially in binaries, i.e. $ f_{\rm bin} = 1.0 $. 
Since $ \sim (60-90)\% $ of OB stars are observed as members of binary systems \citep{md17},
this assumption may lead to overestimate the population size of merging BH--CS binaries by a factor of less than 2.
The primary masses are assumed to obey the initial mass function \citep{kt93},
\begin{eqnarray}
  \xi (M_{1} ) = \left\{
    \begin{array}{ll}
      0                              &\,\, M_{1} \leq 0.1M_\odot \\
      a_{1} M_{1}^{-1.3}  &\,\, 0.1M_\odot < M_{1} \leq 0.5M_\odot \\
     a_{2} M_{1}^{-2.2}  &\,\, 0.5M_\odot < M_{1} \leq1.0M_\odot  \\
     a_{2} M_{1}^{-2.7}   &\,\, 1.0M_\odot <  M_{1} <  \infty.
    \end{array}  \right.,
\end{eqnarray}
where $ a_{1} = 0.29056$ and $ a_{2} = 0.15571$ are the normalized parameters. Thus
\begin{eqnarray}
\Phi (\ln M_{1}) = M_{1}  \xi (M_{1} ).
\end{eqnarray}
The secondary masses are assumed to follow a flat distribution between 0 and $ M_{1}$ \citep{kf07}, 
then 
\begin{eqnarray}
\varphi (\ln M_{2}) = \frac{M_{2}}{M_{1}}.
\end{eqnarray}
The orbital separations are assumed to be uniformly distributed in the logarithm \citep{a83}, thus 
\begin{eqnarray}
\Psi (\ln a)  = k = \rm const.
\end{eqnarray}
The normalization of this distribution yields $ k = 0.12328 $.

\section{Mass transfer stability}

We use the one-dimensional stellar evolution code \textit{MESA} \citep[version 10398,][]{p11,p13,p15,pb18,pb19} to simulate the detailed 
evolution of a large grid of BH binaries with a nondegenerate companion star.  
The BH is treated as a point mass and its binary companion is initially a zero-age main sequence star. The evolutionary models
are computed at solar metallicity ($ Z_{\odot} = 0.02$) and two sub-solar metallicities ($ Z =  $ 0.001 and 0.0001). 
The initial hydrogen mass fraction is assumed
to be $ X= 1-Y-Z  $, where $ Y = 0.24+2Z $ is the helium mass fraction \citep{tp96}. 
For the initial BH binaries, we take the BH masses distributed 
over a range of $ 3-20M_{\odot} $ in logarithmic steps of 0.3, the companion masses over a range of $ 1-100M_{\odot} $ 
in logarithmic steps of 0.05, and the orbital periods over a range of $ 1-10000 $ days in logarithmic steps of 0.1.

In the \textit{MESA} code, convective mixing is accounted for 
by using the mixing-length theory with a default mixing length parameter of $ \alpha = 2 $.  Following \citet{bd11}, 
we include convective core-overshooting with an overshooting parameter of 0.335 pressure scale heights.  
Stellar winds are employed using mass-loss rate prescriptions similar to those in the BPS calculation. 
The wind mass-loss rates of \citet{vink01} are used for 
hot stars and the \citet{nd90} prescription for relatively cool stars with effective temperatures lower than $ 10^{4} $ K. For 
hydrogen-envelope stripped stars ($ X < 0.4 $), we use the reduced rates of \citet{ham95} for helium stars. 
We linearly interpolate the rates between different prescriptions to ensure a smooth transition as described by \citet{bd11}.
We adopt the scheme of \citet{kr90} to calculate the mass-transfer rates $\dot{M}_{\rm tr}$ via RLO. 
Mass accretion onto the BH is limited by the Eddington rate $\dot{M}_{\rm edd}$, and 
the excess matter escapes from the binary system carrying away the specific orbital angular momentum 
of the BH \citep[see also][]{sl20}. 
We simulate the evolution by employing the default timestep options with $ \rm mesh\_delta\_coeff = 1.0  $ and 
$ \rm varcontrol\_target = 10^{-4}$. Each binary evolution track is 
terminated if carbon is depleted in the companion's core or the time steps exceed 30000.

\begin{figure*}[hbtp]
\centering
\includegraphics[width=1.05\textwidth]{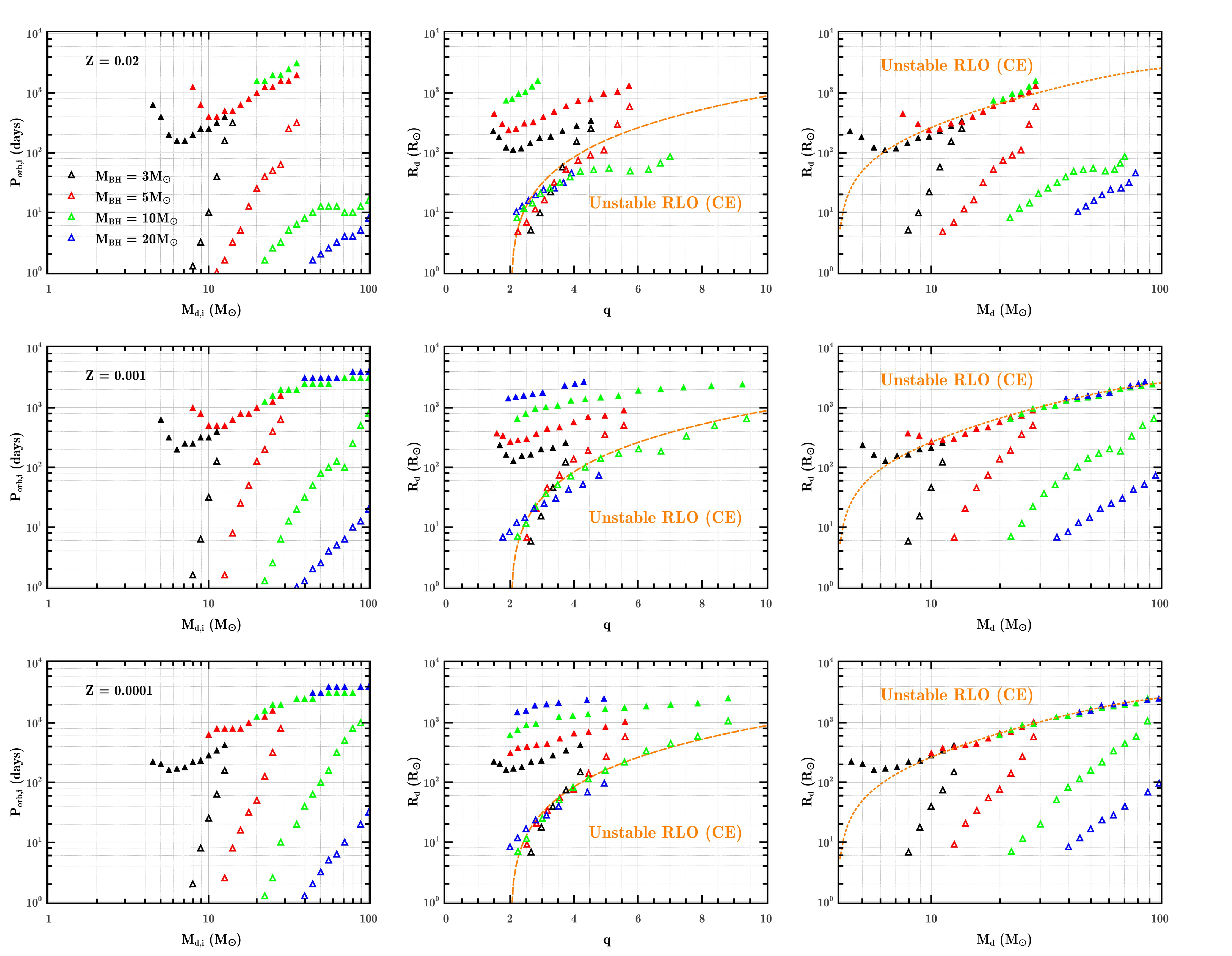}
%\linespread{0.7}
\caption{The parameter spaces for stable and unstable mass transfer in the BH binaries with nondegenerate donors with different 
metallicities. The left panels show the parameter space outlines distributed in the $ M_{\rm d,i} -P_{\rm orb,i}$ diagram, 
while the middle and right panels 
correspond to the cases in the $ q - R_{\rm d} $ and $ M_{\rm d} - R_{\rm d} $ diagrams, respectively. The coloured triangles in each 
panel represent our calculated boundaries for the binaries with different initial BH masses, and the filled and open ones denote
the upper and lower boundaries, respectively. The orange dashed curves roughly fit  
the lower boundaries of the donor radius as a function of the mass ratio, while the orange dotted curves for the upper boundaries 
as a function of the donor mass. 
   \label{figure1}}
\end{figure*}

\citet{pi17} showed that the mass-transferring BH binaries with a massive donor are likely to experience the expansion or
the convective instability, and enter the CE evolution. The
expansion instability occurs when the donor stars are experiencing a period of fast thermal-timescale  
expansion, 
%probably leading to extremely fast mass transfer and dynamical instability. 
while the convection instability takes place if the donor stars have developed a sufficiently deep convective envelope. Usually
the former can be triggered in relatively close BH binaries, while the latter in relatively wide systems. It was suggested by \citet{pi17} that
there exist the smallest radius $ R_{\rm U} $  and the maximum radius $ R_{\rm S} $
for which the convection and expansion instabilities can take place respectively. Mass transfer is stable if the donor radius is between $ R_{\rm U} $ 
and $ R_{\rm S} $. These modified CE criteria 
have been applied to the BPS study on the formation of double CS mergers \citep{ob21}, but the limitation is 
that the calculated outcomes presented by \citet{pi17} are only from a small sample of binary evolution calculations.  

For a BH binary with RLO mass transfer,
the transferred material from the donor star forms an accretion disk around the BH. If the mass transfer proceeds at a very high
rate, the matter will pile up around the BH and presumably form a bloated cloud engulfing a significant
fraction of the accretion disk. Within a specific radius, photons can be trapped in the cloud. This is the so-called trapping radius
that defined by \citet{b79} as 
$R_{\rm trap}  = (R_{\rm sch}/2) (\dot{M}_{\rm tr}/\dot{M}_{\rm edd})$,
where $ R_{\rm sch} $ is the Schwarzschild radius of the BH. We assume that the BH binary is engulfed in a CE if $ R_{\rm trap} $
is larger than the RL radius $ R_{ L_{\rm BH}} $ of the BH, or the 
mass transfer rate exceeds a critical value, $ \dot{M}_{\rm trap}  = 2 \dot{M}_{\rm edd} R_{L_{\rm BH}}/R_{\rm sch} $
\citep{kb99,bk08}. In other conditions, RLO via the $ L_{1} $ point might not be rapid enough to remove the donor's expanding 
envelope, so the donor star extends far beyond to reach the $ L_{2} $ point. Mass loss
through the $ L_{2} $ point can take away a large amount of angular momentum from the binary system and lead to rapid binary orbit 
shrink, probably followed by the CE evolution \citep[e.g.][]{ge20,mf20}. When dealing with the mass transfer stability in 
binaries with convective giant donors, \citet{pi15} showed 
that the binaries can survive the mass transfer even after $ L_{2} $ overflow without starting a CE phase. 
In our simulations, we assume that CE evolution takes place if either of the following conditions is met: (I)  
$ \dot{M}_{\rm tr} >  \dot{M}_{\rm trap}$; (II) $ \dot{M}_{\rm tr} >  0.02M_{\rm d}/P_{\rm orb}$ \citep{pi15} and  
$R_{\rm d} > R_{L_{2}} $ \citep{ge20}. Here $ M_{\rm d}  $ is the donor mass, $ P_{\rm orb} $  the orbital
period of the BH binary, $ R_{\rm d}$ the donor radius, and $ R_{L_{2}} $ the volume-equivalent 
radius of the $ L_{2} $ lobe. When CE starts, we terminate the calculation and
record the relevant information for further analysis.

In Figure 2, we outline our calculated outcomes by showing the parameter space boundaries of the BH binaries 
between stable and unstable mass transfer. The panels from top to bottom correspond to the donor stars 
with different metallicities. The left panels show the boundaries for initial binaries in the  
orbital period $ P_{\rm orb,i} $ vs. donor mass $ M_{\rm d,i} $ diagrams, while the middle and right panels correspond to the 
boundaries for the binaries at the moment of starting RLO in the
mass ratio $ q $ (of the donor to the BH) vs. donor radius $ R_{\rm d} $ and 
donor mass $ M_{\rm d} $ vs. donor radius $ R_{\rm d} $ diagrams, respectively. In each diagram, the coloured triangles 
represent the upper and lower boundaries for dynamically stable mass transfer with different initial BH masses. 
Our results are roughly coincident with those of \citet{pi17}, showing that the binaries with the mass ratios up to $ \sim 10 $ 
can still proceed on a thermal timescale without entering CE evolution \citep[see also][]{sl18,mp21}. In more detail,
our simulations indicate that mass transfer in BH binaries is always  
stable if the mass ratio $ q $ is smaller than the minimal value $ q_{\rm min} $, i.e.
\begin{eqnarray}
q < q_{\rm min}  \sim 1.5-2.0,
\end{eqnarray}
and always unstable if the mass ratio $ q $ is larger than the maximal value $ q_{\rm max} $, i.e.
\begin{eqnarray}
q > q_{\rm max} \sim 2.1 + 0.8 M_{\rm BH},
\end{eqnarray}
where $ M_{\rm BH} $ is the BH mass\footnote{We note that Equation (8) can also be applied to the binaries with an NS accretor, 
where the maximal mass ratio is $ \sim 3-3.5 $ \citep[e.g.,][]{kd00,t00,prp02,sl12,mf20}.}. For the binaries with mass 
ratio between $ q_{\rm min} $ and $ q_{\rm max} $, dynamically unstable mass transfer ensues if the donor radius is either 
less than $ R_{\rm S} $, i.e.
\begin{eqnarray}
R_{\rm d} < R_{\rm S} \sim 6.6-26.1 q + 11.4 q^{2} ,
\end{eqnarray}
or larger than $ R_{\rm U} $, i.e.
\begin{eqnarray}
R_{\rm d} > R_{\rm U} \sim -173.8+45.5M_{\rm d} -0.18M_{\rm d}^{2}. 
\end{eqnarray}
Here all radii and masses are expressed in solar units. We roughly fit $ R_{\rm S}  $ and $R_{\rm U}$ as a function of 
$ q $ (the orange dashed curve) and $ M_{\rm d} $ (the orange dotted curve), respectively. When analysing our recorded data,
we find that Equations (9) and (10) correspond to conditions (I) and (II), respectively. 
%It is indicated that the occurrence of expansion instability is mainly controlled 
%by the mass ratios of the binary system, while the main factor of regulating convective instability is the natures (or the masses) 
%of the donor star. 
We also find that our obtained parameter spaces for stable mass transfer are not strongly dependent on stellar metallicities, 
except for the binaries with donors initially
more massive than $ \sim 40 M_{\odot} $. At solar metallicity, mass transfer in long-period systems with very massive donors are 
always stable since the donor stars have experienced extensive wind mass loss prior to mass transfer \citep{kn21}.
As a consequence, we see in the $ Z=0.02 $ case that the green and blue filled triangles do not appear to show the upper boundaries
for the binaries with $ M_{\rm d,i} \gtrsim 40 M_{\odot}$.
Thus, we adopt Equations (7-10) 
to determinate whether the BH binaries enter the CE evolution in the BPS calculation. There are two exceptions:
for the BH binaries with helium-star donors, we assume they can always avoid CE evolution \citep{tlp15};  for the
BH binaries with WD donors, we assume mass transfer proceeds stably if the mass ratio of the WD to the BH is less than 
0.628 \citep{h02}. When CE evolution is triggered, we allow donors that are crossing the Hertzsprung gap
to survive the CE phase \citep[the ``optimistic" scenario of][]{bk08}.

%\section{Results}

\begin{table*}
\begin{center}
\caption{Predicted occurrence rates for different types of BH$ - $CS binaries.
\label{tbl-1}}
\begin{tabular}{lcccccc}
\\
\hline
Supernova Model     & $ R_{\rm BHWD}$ & $ R_{\rm BHNS}$  & $ R_{\rm BHBH}$   &  $\mathcal{R}_{\rm BHWD}$  
& $ \mathcal{R}_{\rm BHNS} $  & $ \mathcal{R}_{\rm BHBH}$     \\     
   &  ($\rm Myr^{-1} $) & ($\rm Myr^{-1} $) & ($\rm Myr^{-1} $) &  ($\rm Gpc^{-3}yr^{-1} $) &   ($\rm Gpc^{-3}yr^{-1} $)  
   &($\rm Gpc^{-3}yr^{-1} $)   \\
\hline
Rapid   & 11 (0.0) & 8.0 (2.7) & 36 (7.0) & 0.0 [0.0] & 17.4 [0.0] & 42.6 [0.0] \\
Delayed     & 15 (0.29) & 3.6 (1.0) & 19 (6.4) & 6.5 [0.99] & 10.2 [0.68] & 47.2 [0.38] \\
Stochastic     & 95 (3.0) & 33 (5.9) & 150 (17) & 58.6 [0.99] & 71.7 [0.75] & 76.1 [0.28]\\ 
\hline
\end{tabular}
\end{center}
Notes. $ R $ denotes the formation (merger) rate for the BH$ - $CS binaries in the Milky Way. $ \mathcal{R} $ denotes 
the merger rate density [the fraction of systems with BHs being in the mass gap]  for the BH$ - $CS binaries in the local Universe. 
\end{table*}

%When merging BH$ - $CS are formed, the subsequent evolution of the binary systems is controlled
%by gravitational wave radiation \citep{p64}.

\section{Populations of merging BH$ - $CS binaries in the Milky Way}

\begin{figure*}[hbtp]
\centering
\includegraphics[width=0.85\textwidth]{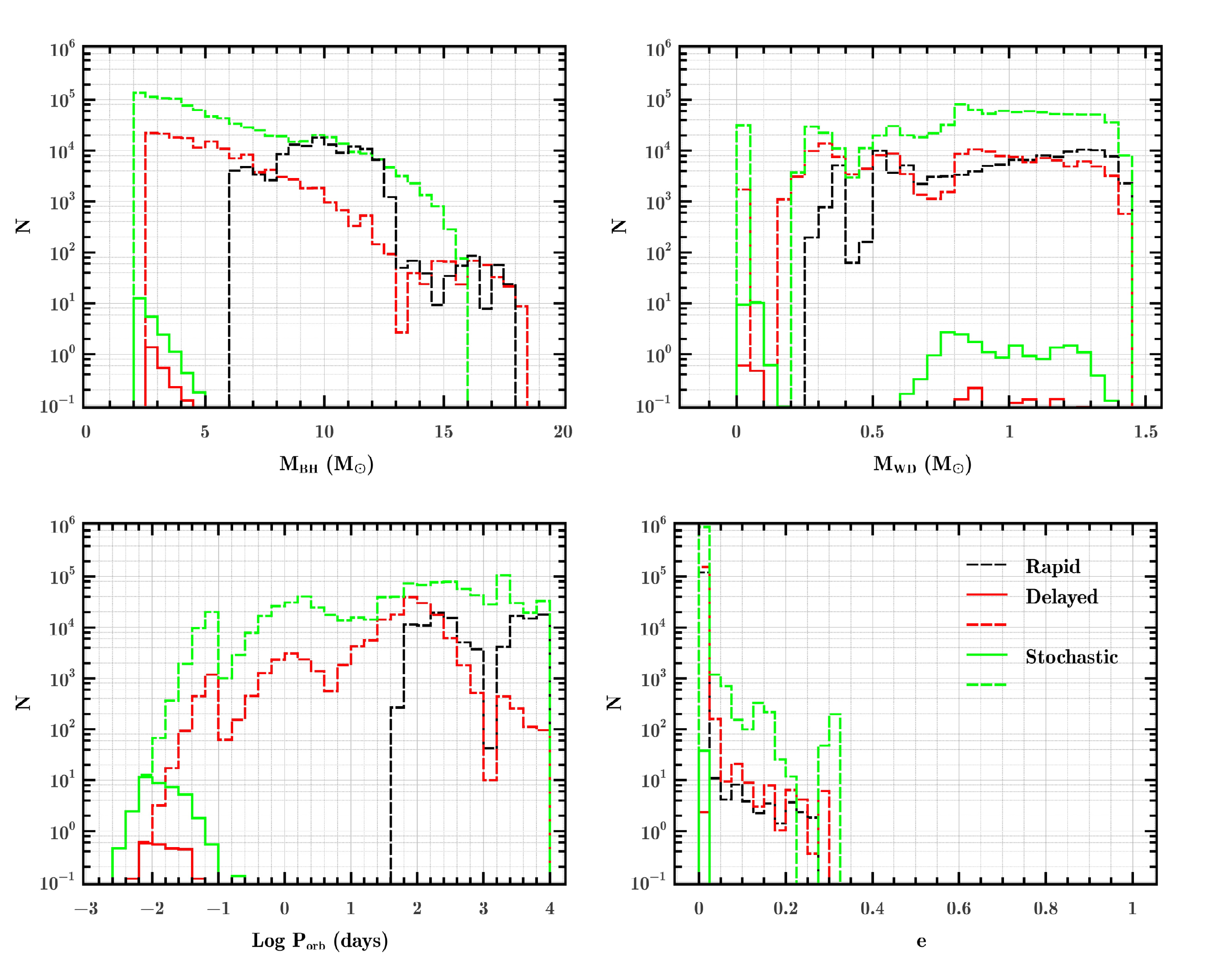}
%\linespread{0.7}
\caption{The solid curves represent the calculated number distributions of detectable BH$ - $WD binaries by LISA 
as a function of the BH mass, the WD mass, the orbital period,
and the eccentricity, assuming continuous star formation rate of $ 3 M_{\odot}\,\rm yr^{-1} $ with solar metallicity over the past 10 Gyr.
The black, red and green curves represent the results using the rapid, delayed and stochastic models, respectively.
The dashed curves correspond to all Galactic BH$ - $WD binaries for comparison. 
   \label{figure1}}
\end{figure*}

Galactic BH$ - $CS binaries are the high-mass analogues of the systems containing only NSs and/or WDs. 
The BH$ - $WD and BH$ - $NS systems
may be discovered through observations of electromagnetic wave signals. On the other hand, close BH$ - $CS binaries
are likely to be observed by future space-based observatories due to the detection of GW signals. 
We will evaluate the population properties of  close BH$ - $CS binaries that can be identified by LISA. 

The angle-averaged signal-to-noise ratio for a BH$ - $CS system that can be detected over an 
observation time $ T $  is given by \citep{ok09}
\begin{eqnarray}
\left\langle ({\rm S/N})^{2} \right\rangle  = \sum_{n} \int \left[ \frac{h_{c,n}(f_{n})}{h_{N}(f_{n})} \right]^{2} d \ln f_{n},
\end{eqnarray}
where $ n $ labels the harmonics at frequency $ f_{n} \simeq n f_{\rm orb}$, 
\begin{eqnarray}
h_{c,n}^{2}= \frac{1}{(\pi d_{\rm L})^{2}}\left( \frac{2 G}{c^{3}} \frac{\dot{E}_{n}}{\dot{f}_{n}}\right) 
\end{eqnarray}
is the characteristic strain at the $ n $th harmonic \citep{bc04}, and $ h_{N}(f_{n}) $ is the 
characteristic LISA noise, including a contribution from unresolved Galactic binaries, for which we take from \citet{rcl19}.
Here $ T$ is taken to be $ 4\rm\, yr $ for the LISA mission duration,  $ f_{\rm orb} $ is the orbital frequency of the binary, 
$ d_{\rm L} $ is the luminosity distance to the source,  
$ G $ is the constant of gravity, and $ c $ is the speed of light in vacuum. In Equation (12), $ \dot{E}_{n} $ is the 
derivative of the energy radiated in GWs at frequency $ f_{n} $, which to lowest order is given as \citep{pm63}
\begin{eqnarray}
\dot{E}_{n} = \frac{32}{5} \frac{G^{7/3}}{c^{5}} (2 \pi f_{\rm orb} M_{\rm chirp})^{10/3} g(n,e),
\end{eqnarray}
where $ M_{\rm chirp} = (M_{\rm BH}M_{\rm CS})^{3/5}(M_{\rm BH}+M_{\rm CS})^{-1/5} $ is the chirp mass, $ M_{\rm BH} $
and $ M_{\rm CS} $ are the component masses of the BH$ - $CS binary, and $g(n,e)$ is a function of the orbital eccentricity $ e $
\citep[from][]{pm63}. To the leading quadrupole order, the term $ \dot{f}_{n} $ is 
\begin{eqnarray}
\dot{f}_{n} = n \frac{48}{5 \pi} \frac{(GM_{\rm chirp})^{5/3}}{c^{5}}(2 \pi f_{\rm orb})^{11/3} F(e),
\end{eqnarray}
where $ F(e) = [1+(73/24)e^{2} +(37/96)e^{4}]/(1-e^{2})^{7/2} $. 
Note that a source is effectively less detectable the slower the GW frequency changes over the instrumental lifetime. 
This effect is taken into account by reducing the characteristic strain $ h_{c,n} $ by a factor of the square root of
$ \min [ 1, \dot{f}_{n}(T/f_{n})] $ \citep[see e.g.,][]{kr19}.

The peak frequency of GW emission for eccentric binaries is given by 
\begin{eqnarray}
f_{\rm GW} = n_{p} f_{\rm orb},
\end{eqnarray}
where $ n_{p}  \simeq 2 (1+e)^{1.1954}/(1-e^{2})^{1.5}$ \citep{w03}. Obviously, $ n_{p}=2 $ for circular binaries.
Considering that the majority of LISA-visible binaries have mild eccentricities, the GW power is sharply peaked at
the peak frequency \citep{pm63}. We follow \citet{bs20} to simplify the signal-to-noise ratio  as
\begin{eqnarray}
\left\langle ({\rm S/N}) \right\rangle \simeq  \min  \left[ 1, \sqrt{\dot{f}_{n}(T/f_{n})} \right] \frac{h_{c,n}(f_{\rm n})}{h_{N}(f_{\rm n})},
\end{eqnarray} 
with $ n = n_{p} $ and $ f_{n} = f_{\rm GW}$.
We assume that a BH$ - $CS binary can be identified by LISA if the signal-to-noise ratio is larger than 5 \citep[e.g.,][]{lg18,bs20}.
To obtain 
the $ d_{\rm L} $ distribution of the BH$ - $CS binaries in the Milky Way, we assume that the binaries are  
uniformly distributed on a flat disc with a radius 15 kpc\footnote{Although this assumption is simple, it can roughly reflect the spatial
distribution of detectable LISA binaries \citep[see Figure 1 of][]{lmv20}.}
and the Sun with respect to the Galactic Center has the distance of 8 kpc \citep{fw97}.   
When synthesizing the BH$ - $CS binary population, we adopt continuous star formation at a rate
$ {\rm SFR} = 3 M_{\odot}\,\rm yr^{-1} $ \citep{sb78,dh06,rw10} with solar metallicity over a period of 10 Gyr. 
In Table 1, we present the formation and merger rates of Galactic BH$ - $CS binaries in our adopted three supernova models.
Under all the models, we obtain that Galactic BH$ - $WD, BH$ - $NS and BH$ - $BH binaries have the formation rates 
$\sim  11-95 \rm \,Myr^{-1} $, $  4-33 \rm\, Myr^{-1} $ and 
$ 19-150 \rm\, Myr^{-1} $, respectively, and the merger rates 
$ \sim 0-3 \rm \,Myr^{-1}$, $  1-6 \rm \,Myr^{-1}$ and $ 6-17 \rm \,Myr^{-1}$, respectively.
It can be estimated that the total number of Galactic BH$ - $CS binaries formed over the past 10 Gyr is of the
order $ 10^{5}-10^{6} $. 
%\citet{lmv20} proposed two extreme scenarios for the spatial distribution when investigating Galactic NS$ - $NS 
%systems as potential LISA sources. The one assumed negligible kicks at NS formation, so that the distribution of
%the NS$ - $NS binaries is limited by the disc gravitational potential. The other one assumed large magnitudes of
%kick velocities, under which the binaries follow the mass distribution of the dark matter hole. 

\begin{figure}[hbtp]
\centering
\includegraphics[width=0.5\textwidth]{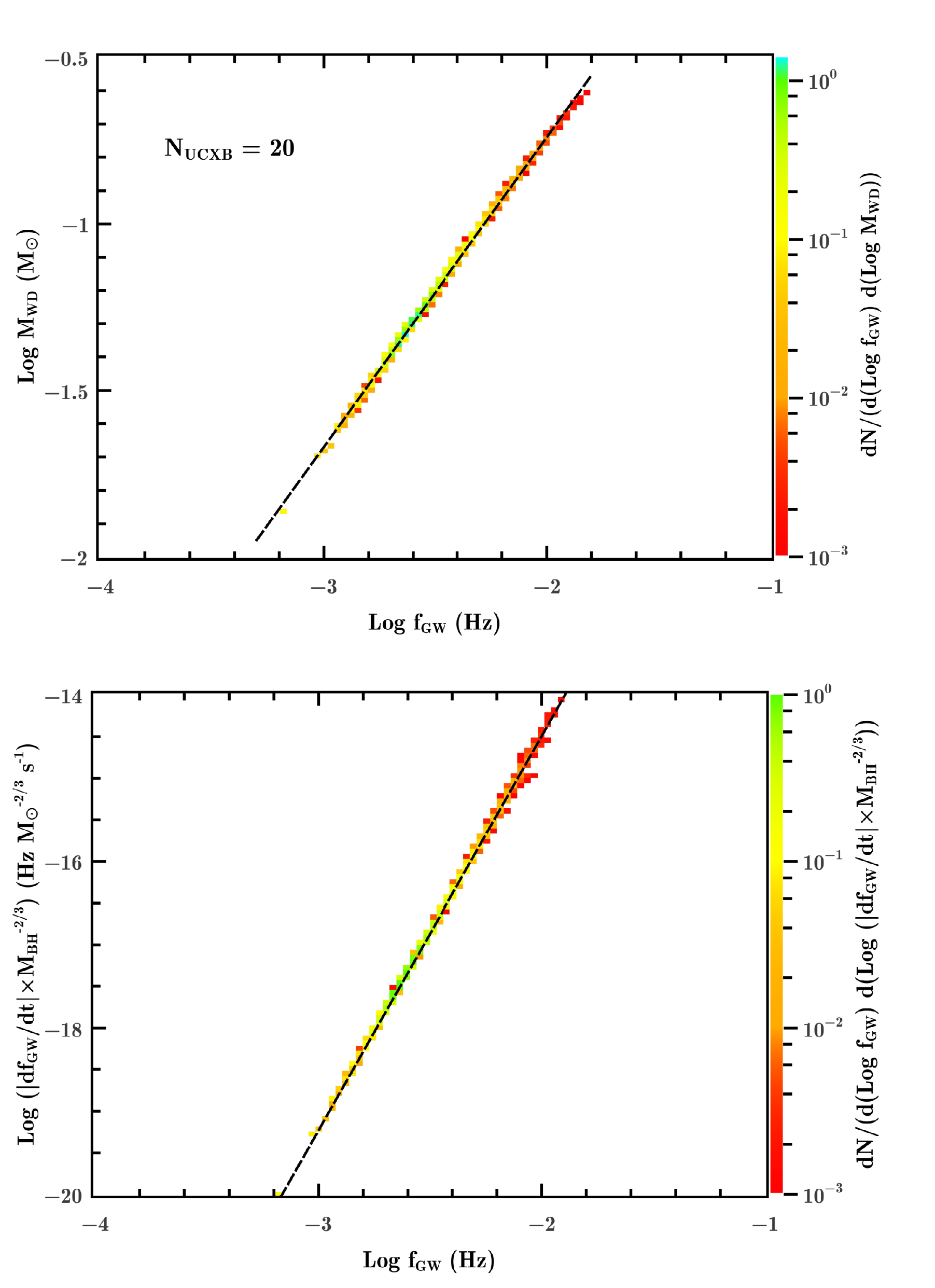}
%\linespread{0.7}
\caption{Predicted number distributions of Galactic BH$ - $WD UCXBs as potential LISA sources in 
the $ M_{\rm WD} -  f_{\rm GW}  $ and  $ |\dot{f}_{\rm GW}| M_{\rm BH}^{-2/3} -  f_{\rm GW}$ planes under 
the stochastic model. The colors in each pixel are scaled according to the numbers of the binary systems.
We fit the data with linear relationships of  $ \log M_{\rm WD} = 1.12+0.93\log  f_{\rm GW}  $ and  $\log (|\dot{f}_{\rm GW}| M_{\rm BH}^{-2/3}) = -5.04 + 4.73 \log f_{\rm GW}$, as plotted in the two dashed lines. 
   \label{figure1}}
\end{figure}

\begin{figure*}[hbtp]
\centering
\includegraphics[width=0.85\textwidth]{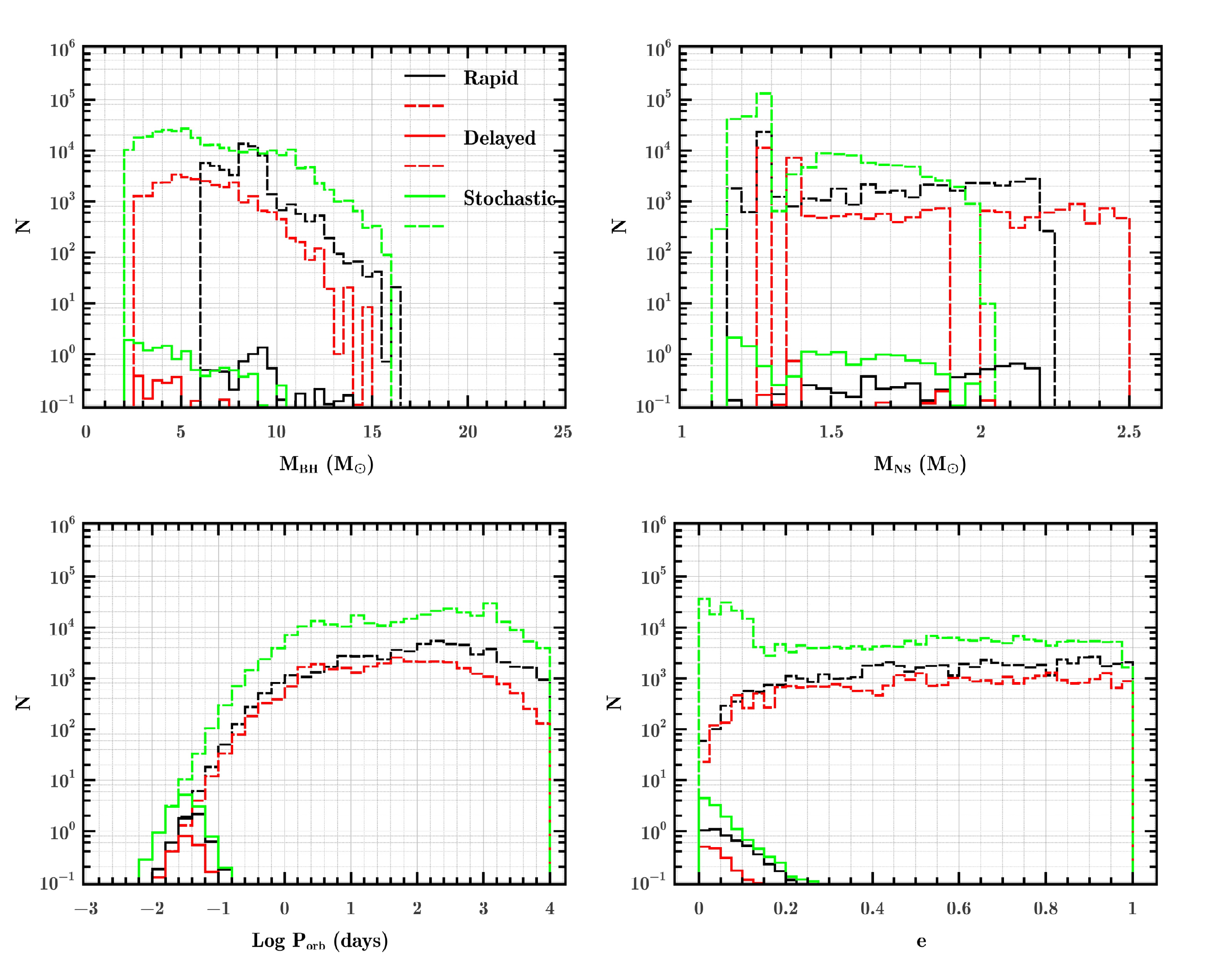}
%\linespread{0.7}
\caption{Similar to Figure 3, but for BH$ - $NS binaries. 
   \label{figure1}}
\end{figure*}

\begin{figure*}[hbtp]
\centering
\includegraphics[width=0.85\textwidth]{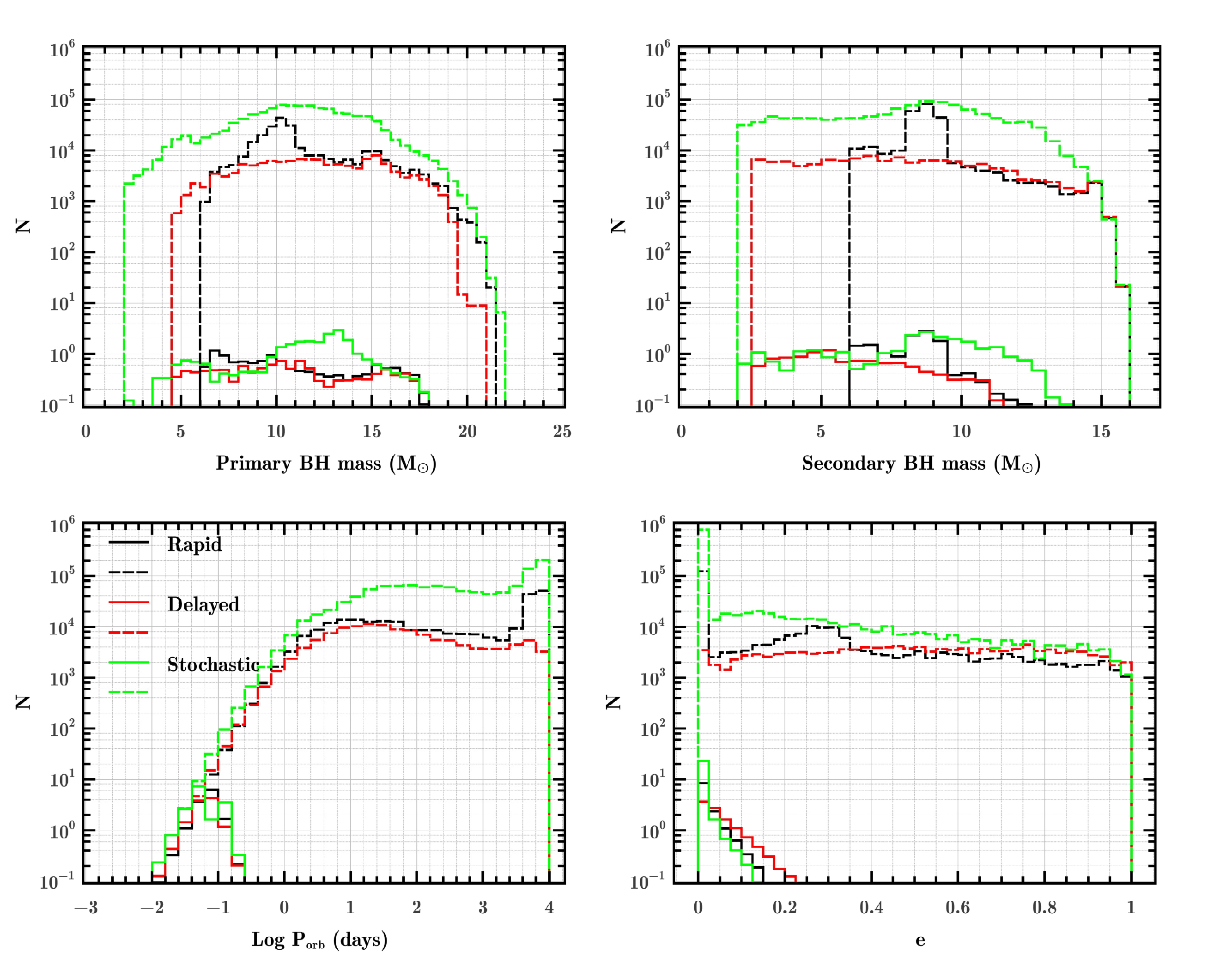}
%\linespread{0.7}
\caption{Similar to Figure 3, but for BH$ - $BH binaries. 
   \label{figure1}}
\end{figure*}

\subsection{The case of BH$ - $WD systems}

Figure 3 presents the histogram distributions for the calculated number of BH$ - $WD binaries as a function of
the BH mass, the WD mass, the orbital period and the eccentricity under assumption of different models. The solid and
dashed curves correspond to detectable BH$ - $WD binaries by LISA and all the Galactic systems (for comparison), respectively.  Compared with 
the delayed and the stochastic models, it is clearly seen that there are no close BH$ - $WD binaries formed in the rapid model. 
The reason is that the rapid model can only produce BHs massive than $ 5 M_{\odot} $, so mass transfer from an intermediate-mass donor
star is always stable, which causes the resulting BH$ - $WD systems to have orbital periods longer than 30~days. 
In comparison, both the delayed and the stochastic models allow the formation 
of mass-gap BHs, CE phases can take place during the progenitor evolution involving a $ \sim 2 - 5M_{\odot} $ BH and 
a $ \sim 6-10M_{\odot} $ donor (see Section 3), probably resulting in the creation of close BH$ - $WD systems. There are 
$ \sim 2 $ and $ \sim 38 $ detectable LISA binaries generated in the delayed and stochastic models, respectively. 
We find that all detectable BH$ - $WD binaries by LISA host BHs within the mass gap and the BH mass distribution has
a peak at $ \sim 2-3 M_{\odot} $.  According to the distribution of the WD masses, the LISA sources can be classified into
two groups with $ M_{\rm WD} \lesssim  0.1 M_{\odot}$ and $M_{\rm WD} \sim 0.6-1.4M_{\odot} $. 
All the BH$ - $WD binaries have circular orbits with periods of $ \lesssim 0.1 $ day.

The LISA systems with $M_{\rm WD} \sim 0.6-1.4M_{\odot} $ are detached binaries.
The orbital shrink due to GW radiation may lead the originally detached binaries to begin RLO, evolving to be 
UCXBs. Mass transfer proceeds rapidly
in the binaries with WD donors massive than $ \sim 0.1M_{\odot} $, so such mass-transferring systems have negligible contribution to the 
whole LISA binary population. Subsequently, mass transfer may settle into an equilibrium state when the response of the RL radius matches 
the one of the WD radius \citep[see also][]{st21}. Besides the detached systems with a $ \sim 0.6-1.4M_{\odot} $ WD, 
the LISA sources may also be observed as UCXBs with a $ \lesssim  0.1 M_{\odot} $ WD donor around a BH accretor.
Note that these UCXBs should appear as expanding rather than merging systems since mass transfer tends to widen the binary orbits. 
Based on the delayed (stochastic) model, we estimate that there are about $ \sim 1 $ ($ \sim 20 $) BH$ - $WD binaries that may be 
observed via both electromagnetic and GW signals. 
For the systems with WDs more massive than $ \sim 1 M_{\odot} $, the binary orbits at the onset of RLO are so compact that
the orbital decay due to GW radiation can 
overcome the orbital expansion due to mass transfer, finally leading the binaries to merge.

\begin{figure*}[hbtp]
\centering
\includegraphics[width=1.05\textwidth]{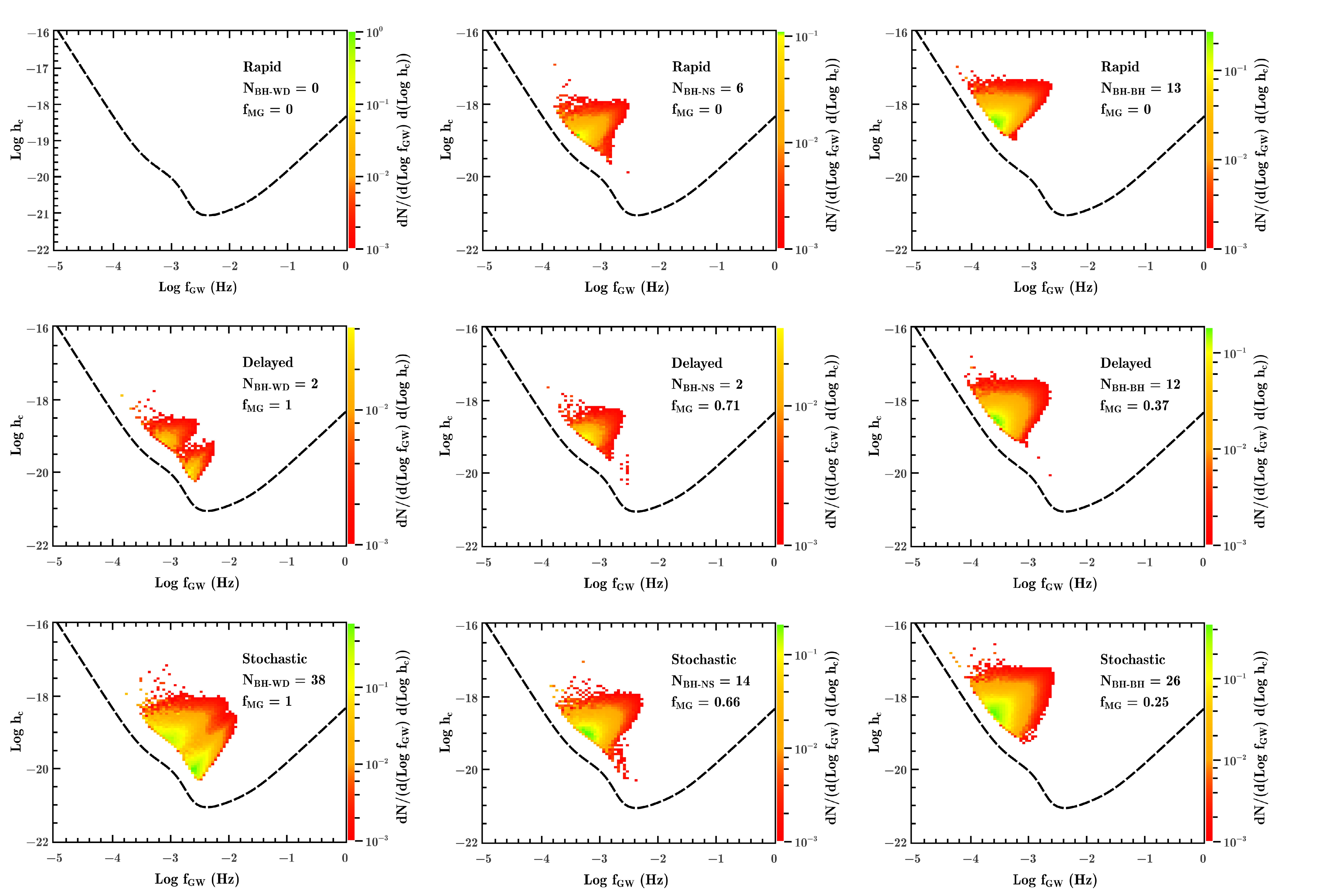}
%\linespread{0.7}
\caption{Predicted number distributions of observable GW sources for Galactic BH$ - $CS binaries in the LISA frequency band. 
The rapid, delayed and stochastic models correspond to the different panels from top to bottom. In each panel, we have labelled the  
number of corresponding GW sources and the fraction $ f_{\rm MG} $ of systems hosting (at least) a mass-gap BH. The dashed curve
denotes the LISA sensitivity curve fitted by \citet{rcl19}.  The colors in each pixel are scaled according to the corresponding numbers of the 
GW sources.
   \label{figure1}}
\end{figure*}

Using a semianalytical approach, \citet{st21} identified two universal relationships for mass-transferring BH$ - $WD binaries:
$ M_{\rm WD} $ vs.  $  f_{\rm GW}  $ and $ \dot{f}_{\rm GW} M_{\rm BH}^{-2/3} $ vs. $ f_{\rm GW}$. 
The mass$ - $radius relation of WDs and the condition of 
$ M_{\rm WD} \ll M_{\rm BH}$ are implicitly used to derive the above relationships. In Figure~4, we show the number distributions
of Galactic BH$ - $WD UCXBs that are likely to be detected by LISA in the $ M_{\rm WD} -  f_{\rm GW}  $ (top) and 
$ |\dot{f}_{\rm GW}| M_{\rm BH}^{-2/3} -  f_{\rm GW}$ (bottom) planes under the stochastic model. Each panel
contains a $ 100\times 100 $ matrix element for the corresponding parameters. The color in each pixel denotes the number of 
LISA UCXBs in the matrix element by accumulating the product of their birthrates of the systems passing 
through it with the time duration. Our simulated outcomes confirm the relationships 
proposed by \citet{st21}. The GW frequency of BH$ - $WD UCXBs can cover the range of $ \sim 1-10 $ mHz. Importantly, 
these relationships may be applied to disentangle the component masses of LISA BH$ - $WD UCXBs. 
It is possible that they are also suitable for the systems with a WD donor when the accretor is 
an NS \citep{ttm18,clw20,wcl21} or even another massive WD \citep{ny04,kbl17}, which is beyond the scope of this paper.
%in the LISA sources with a WD donor when the CS
%is a BH,  even an NS or another massive WD. 

\subsection{The case of BH$ - $NS systems}

In Figure 5 we plot the histogram number distributions of observable BH$ - $NS systems by LISA as a function of the binary
parameters in the rapid (black curves), delayed (red curves) and stochastic (green curves) models. The dashed curves
correspond to all BH$ - $NS binaries in the Milky Way for comparison. The rapid model predicts that the BH
masses of LISA binaries are distributed with a peak at $ \sim 8-9 M_{\odot} $, while the delayed and stochastic models favor 
creating the binaries with BHs being in the mass gap. The reason is that the 
binaries with light BH progenitors (see Figure~1) tend to have high formation rates due to the initial mass function. Besides, in the rapid model, 
the systems are more likely to avoid disruption where BHs are formed via direct collapse without any kick \citep{fb12}.
There is a common feature that quite a fraction of 
Galactic BH$ - $NS binaries contain a $ \sim 1.3M_{\odot} $ NS in all models, corresponding to the NSs originating
from electron-capture supernovae. For the LISA sources, it can be seen that the NS masses have a more flat distribution
in each model. All models predict that the orbital periods of Galactic BH$ - $NS binaries are mainly distributed in a broad range of 
$ \sim 1-10000 $ days, while 
only $ \sim 2-14 $ of systems can appear as the LISA sources with orbital periods 
less than $ \sim 0.1 $ day. The rapid and delayed
models tend to produce the BH$ - $NS binaries with relatively large eccentricities, while the stochastic model favor 
producing the systems with nearly circular orbits since most NSs are formed with low kicks.
As a consequence of the orbital decay via GW radiation, merging BH$ - $NS binaries that can be observed by LISA have relatively
low eccentricities of $ \lesssim 0.2 $. 

NSs are likely to be observed as radio pulsars if they are still active and beamed towards the Earth.  
Among all BH$ - $NS binaries, a very small fraction of them are expected to be identified 
due to the detection of radio pulsars. Based on the formation rates ($ \sim 4-33 \rm\, Myr^{-1} $) of Galactic BH$ - $NS systems 
in our adopted three models, we roughly estimate that there are $ \sim 20-200 $ BH binaries containing a radio pulsar 
in the Milky Way, assuming a transformation factor
of  $\sim 6 \rm Myr $ between the formation rate and the binary number \citep[see Table 1 of][]{sl18}. 
%It is estimated that only $ \sim 10^{-4}-10^{-3} $ of all Galactic 
%BH$ - $NS systems may be observed due to the detection of GW signals or radio pulsars. 

\subsection{The case of BH$ - $BH systems}

In Figure 6 we present similar histogram diagrams for calculated number distributions (solid curves) 
of LISA BH$ - $BH systems under
different supernova models. For comparison, the dashed curves denote all Galactic BH$ - $BH binaries. 
The primary BHs tend to be heavier than the secondary ones, and have the maximal mass of  
$ \sim 18 M_{\odot} $ in LISA BH$ - $BH binaries ($ \sim 22 M_{\odot} $ in all the Galactic binaries). 
The rapid model predicts that the secondary BHs of LISA binaries have the mass distribution
in the range of $ \sim 6-13M_{\odot} $ with a 
peak $ \sim 8-9M_{\odot} $ which is similar to the case for the BH masses of LISA BH$ - $NS systems, while the
other two models anticipate that a 
significant fraction of secondary BHs have masses within the mass gap. As most natal
BHs are imparted by negligible kicks in the rapid and stochastic models, 
a large part of Galactic BH$ - $BH binaries have long periods of $ \gtrsim 3000 $ days in nearly circular orbits. 
In the delayed model, the orbital periods are mainly distributed in the 
range of $ 1-10000 $ days with a peak $ \sim 10 $ days, and the orbital eccentricities have a flat distribution between $ 0-1 $. 
Our calculations show that there are $ \sim 12-26 $ BH$ - $BH binaries detectable by LISA. Compared with the LISA
BH binaries with an NS or a WD companion, these BH$ - $BH systems possess longer orbital periods up to $ \sim 0.3 $ day.
Also, LISA BH$ - $BH binaries are expected to have relatively low eccentricities of $ \lesssim 0.2 $. 

\subsection{GW detection of BH$ - $CS systems}

Figure 7 shows predicted number distributions of detectable BH$ - $CS binaries by LISA in the $ h_{\rm c}-f_{\rm GW} $
plane for the rapid, delayed and stochastic models (from top to bottom panels).  The left, middle and right panels correspond
to the CS companion being a WD, an NS and a BH, respectively. In each panel, we have labelled the predicted number of the
corresponding GW sources and the fraction $ f_{\rm MG} $ of systems hosting BHs within the mass gap. 
In the rapid
model, there are no mass-gap BHs produced, so $ f_{\rm MG} $ are always zero for all three types of BH$ - $CS binaries. 
Both the delayed and the stochastic models predict that all BH$ - $WD binaries detectable by LISA possess a mass-gap
BH, i.e. $ f_{\rm MG} = 1.0 $, and that $ f_{\rm MG} $ decreases to $ \sim 0.7 $ and $ \sim 0.3 $ for 
BH$ - $NS and BH$ - $BH binaries, respectively. In the rapid model, all BH$ - $WD binaries have wide orbits and
do not appear to be LISA sources. In the delayed and the stochastic models, BH$ - $WD systems 
cover two regions that correspond to detached systems (with relatively high $ h_{\rm c} $ values) and 
 UCXBs (with relatively low $ h_{\rm c} $ values). 

\begin{figure}[hbtp]
\centering
\includegraphics[width=0.4\textwidth]{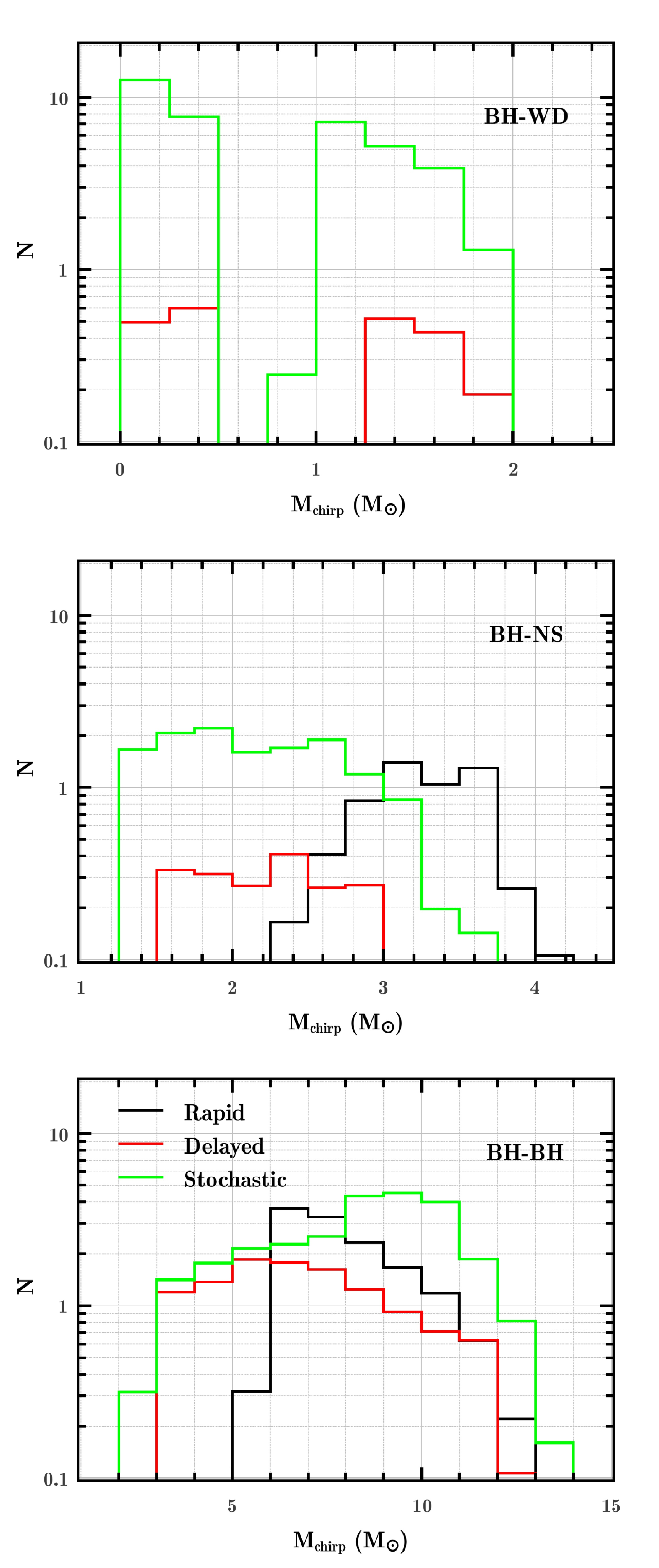}
%\linespread{0.7}
\caption{Calculated number distributions of LISA BH$ - $CS binaries as a function of the chirp mass in all our adopted models. 
The top, middle and bottom panels correspond to the CS companions being WDs, NSs and BHs, respectively.
   \label{figure1}}
\end{figure}

In Figure 8 we show the calculated number distributions of detectable BH$ - $CS binaries by LISA for all our adopted models, as 
a function of the chirp mass. In both the
delayed and the stochastic models, LISA BH$ - $WD binaries can be divided into detached systems with 
$ M_{\rm chirp} \sim 0.8-2M_{\odot}$ and UCXBs with $ M_{\rm chirp} \lesssim 0.5M_{\odot}$. For BH$ - $NS binaries,
the chirp mass distributions can cover the range of $ \sim 1.2-4M_{\odot} $, 
with an overlap between $\sim 2.2-3M_{\odot} $ between the three models.
LISA BH$ - $BH binaries have chirp masses mainly distributed in the range of $ \sim 3-12M_{\odot} $ for the
delayed and stochastic models, and $ \sim 5-12M_{\odot} $ for the rapid model. 
Based on the chirp mass distributions
for different types of BH$ - $CS binaries, LISA may distinguish BH$ - $BH systems
 if $ M_{\rm chirp} \gtrsim 4M_{\odot}$ and 
BH$ - $WD systems if $ M_{\rm chirp} \lesssim 1M_{\odot}$ from BH$ - $NS binaries. However, the identification of BH$ - $CS 
binaries  solely from GW observations is difficult since they can be confused with other types of GW sources such as 
NS$ - $NS systems and NS$ - $WD systems \citep[see also][for the discussion on the identification of BH--BH binaries]{slp20}.

\section{Populations of BH$ - $CS mergers in the local Universe}

To estimate the merger rate density $ \mathcal{R}_{\rm BHCS} $ of the BH$ - $CS binaries in the local Universe, we have evolved a large number
($2 \times 10^{7} $) of the primordial binaries for each model. The metallicity Z for every primordial binary is randomly taken 
in the logarithmic space between 0.0001 and 0.02. 
After simulations, we record relevant information for the BH$ - $CS binaries that can merge within
the Hubble time, including the CS types, the component masses and the delay time $ t_{\rm delay} $ (from the formation of the 
primordial binaries to the merger of the BH$ - $CS binaries). 
For each BH$ - $CS merger, the look back time of the merger can be
estimated as 
\begin{eqnarray}
t_{\rm merg} = t_{\rm lb}-t_{\rm delay},
\end{eqnarray}
where $ t_{\rm lb} $ is the look back time for (binary) stars formed at redshift $ z $,
\begin{eqnarray}
t_{\rm lb}(z) = \frac{1}{H_{0}} \int_{0}^{z} \frac{1}{(1+z)[\Omega_{\rm M}(1+z)^{3}+\Omega_{\Lambda}]^{1/2}} \rm{d}z,
\end{eqnarray} 
where  $ H_{0} $, $ \Omega_{\rm M} $ and $ \Omega_{\Lambda} $ are the cosmological parameters for which we 
adopt the values from \citet{planck16}. Following \citet{mg17},
we consider the BH$ - $CS mergers with 
$ t_{\rm merg} \geq 0 $, excluding the systems that will merge in the future. 
And, we only include the mergers in the local Universe (defined as
$ z\leq 0.1 $) using the condition of $ t_{\rm merg} \leq t_{\rm lb}(z=0.1) $. 
We further divide the recorded binaries into 20 logarithmically spaced metallicity bins between $ Z = 0.0001-0.02 $ for each model. 
The metallicity for stars at a given redshift is computed as $ \log Z(z)/Z_{\odot} = -0.19z $ if $ z \leq1.5 $ and
$ \log Z(z)/Z_{\odot} = -0.22z $ if $ z > 1.5 $ \citep{rw12}. For $ Z < 0.0001 $ or $ Z > 0.02 $, we instead use the recorded information of
the systems with $ Z = 0.0001 $ or $ Z = 0.02 $. In each metallicity bin, the parameter distribution of the primordial binaries has been 
normalized to unity. 
Similar to the treatment of \citet{gm18}, we use the following analytic equation to calculate $ \mathcal{R}_{\rm BHCS} $ as \citep[see also][]{sm19}

\begin{equation}
 \mathcal{R}_{\rm BHCS} = \frac{1}{t_{\rm lb}(z= 0.1)} \sum^{0.1}_{z=15} t_{\rm lb}(z) \left( \frac{f_{\rm bin}}{2} \right) 
 \left( \frac{{\mathcal{SFR}}(z)}{M_{*}} \right) W_{\rm b},
\end{equation}
where $ \mathcal{SFR}(z) $ is  the cosmic star-formation rate density as a function of 
redshift  for which we use the fitted formula given by \citet{md14},

\begin{equation}
\mathcal{SFR}(z) = 0.015 \frac{(1+z)^{2.7}}{1+[(1+z)/2.9]^{5.6}} M_{\odot} \rm Mpc^{-3}  yr^{-1}.
\end{equation}

\subsection{The BH--CS mergers with(out) mass-gap BHs}

In Table 1, we show the predicted local merger rate densities of different types of BH--CS binaries and
the fraction $ f_{\rm MG} $ of the merging binaries that host mass-gap BHs for all our adopted models. 
The delayed and stochastic models predict 
$ \mathcal{R}_{\rm BHWD} \sim 6.5 \rm\, Gpc^{-3}yr^{-1}$ and $ 58.6 \rm\, Gpc^{-3}yr^{-1} $, respectively. 
The local merger rate densities of BH--NS binaries are
in the range of $ \sim 10.2-71.7 \rm\, Gpc^{-3}yr^{-1} $, consistent with the inferred upper limit of 
$ 610 \rm\, Gpc^{-3}yr^{-1}$ from the LIGO/Virgo data\footnote{More recently, \citet{aaa21}
reported the detection of two BH--NS mergers (GW200105 and GW200115) and inferred the merger rate density of 
$ 45^{+75}_{-33} \rm\, Gpc^{-3}yr^{-1}  $ if assuming they are representative of the BH--NS population.
In this case, our calculated results can still match the observations.}  \citep{ab19}. For BH--BH binaries,
the merger rate densities are in the range of $ \sim 42.6-76.1 \rm\, Gpc^{-3}yr^{-1} $,
slightly larger than the one $ 23.9^{+14.3}_{-8.6} \rm\, Gpc^{-3}yr^{-1} $ given by LIGO/Virgo observations
\citep{ab20b}. Our obtained rates are obviously subject to many uncertainties such as the assumptions on
supernova kicks, CE ejection efficiencies, stellar winds, and initial parameter distribution of the primordial binaries. 
For example, the calculated $ \mathcal{R}_{\rm BHBH} $ may better match 
observations if increasing the magnitude of the kick velocities for BHs. 
Compared with BH--CS mergers in the rapid model, we estimate that $ \sim 99\% $ ($ \sim 99\% $)
of BH--WD mergers,  $ \sim 68\% $ ($ \sim 75\% $) of BH--NS mergers, and $ \sim 38\% $ ($ \sim 28\% $) of BH--BH mergers
host mass-gap BHs in the delayed (stochastic) model. We can see that these fractions
are very close to those for the corresponding type of LISA binaries in the Milky Way. 

\begin{figure}[hbtp]
\centering
\includegraphics[width=0.4\textwidth]{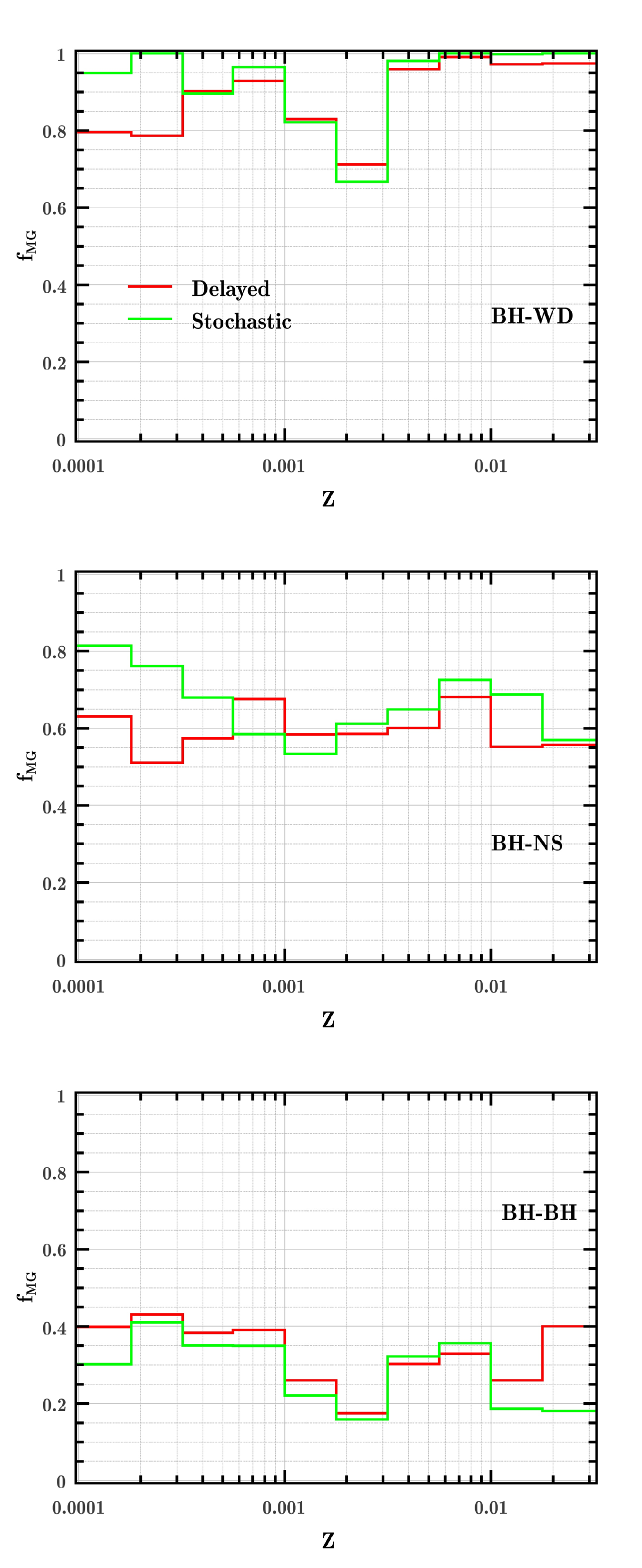}
%\linespread{0.7}
\caption{The fractions $ f_{\rm MG} $ of the mergers with mass-gap BHs among
all mergers for different types of BH--CS systems as a function of the metallicity. The red and green
curves correspond to the delayed and the stochastic models, respectively. The top, middle
and bottom panels correspond to the CS being a WD, an NS and a BH, respectively.
   \label{figure1}}
\end{figure}

Figure 9 shows the fractions $ f_{\rm MG} $ of the BH--CS mergers with mass-gap BHs among all mergers as a function of 
the metallicity. 
%We assume the same star formation rate within the Hubble time for each metallicity bin.
%The rapid model cannot produce mass-gap BHs, i.e. $ f_{\rm MG}  = 0$. 
We find that $ f_{\rm MG} $ is not strongly dependent on metallicity. 
Both the delayed and the stochastic models predict that
$f_{\rm MG} \sim 0.7-1.0$ for BH--WD mergers, $ f_{\rm MG} \sim 0.5-0.8$ for BH--NS mergers, and 
$f_{\rm MG}  \sim 0.2-0.4$ for BH--BH mergers. 

\begin{figure}[hbtp]
\centering
\includegraphics[width=0.4\textwidth]{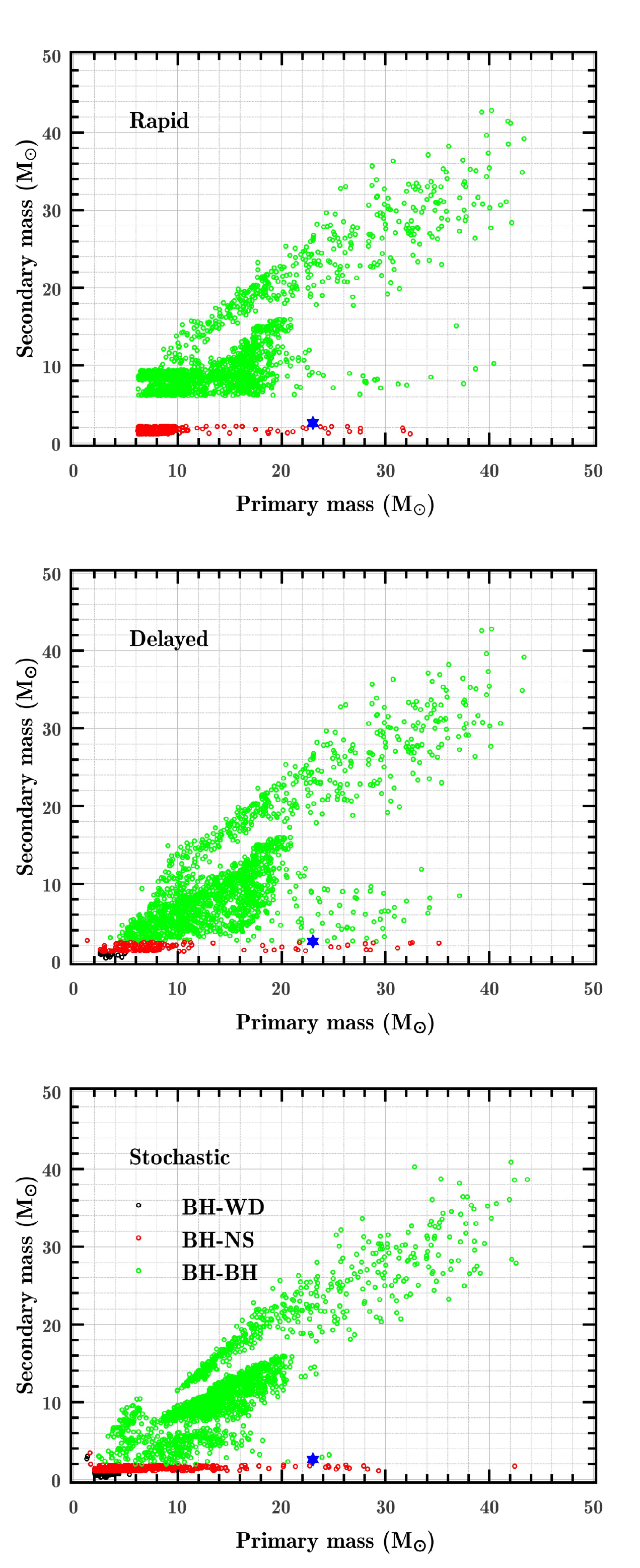}
%\linespread{0.7}
\caption{Component masses of the BH--CS mergers in the local Universe for our adopted three models. The black, red and
green circles correspond to the CSs being WDs, NSs and BHs, respectively. The position of GW190814 is marked with a blue star. 
   \label{figure1}}
\end{figure}

Figure~10 shows the component masses of the BH$ - $CS systems merged in the local Universe for all our adopted models. The black, red
and green circles correspond to the CS companions being WDs, NSs and BHs, respectively. The blue star marks the position of 
GW190814. We can see that GW190814 cannot form in the rapid model if its less-massive component 
is a mass-gap BH. Predictably, some BH--BH binaries can match the component masses of GW190814-like systems in both
the delayed and stochastic models. The formation of GW190814-like sources has been explored by \citet{zs20}, who suggest that
the predicted rate of such mergers is in tension with the empirical LIGO/Virgo rate of other CS pair mergers if only involving isolated 
binary evolution channel. 

Based on our calculated outcomes, it is possible to form BH$ - $NS systems where the NS forms first in the
delayed and stochastic models (see Figure~10 for the binaries with primary mass $ \sim1-2M_{\odot} $), 
but the merger rate of these binaries is $ 1-2 $ orders of magnitude lower than that of the systems
where the BH forms first. In the stochastic model, even a WD can form first in BH$ - $WD systems. Mass transfer efficiency during
the primordial binary evolution is a vital factor to determinate whether the NS/WD can be formed before the BH \citep{spn04}. Since we adopt the
rotation-dependent mass transfer mode in our calculations, the relatively low mass-transfer efficiency makes it difficult to form 
the NS/WD before the BH. 

\subsection{Formation channels of BH--CS mergers}

From the evolutionary point of view, BH--CS systems form either from stable mass transfer channel or the CE channel. 
Considering the evolution of the BH binaries with nondegenerate donors experienced either a stable mass transfer phase or a CE phase, we 
identify two channels for the formation of BH$ - $CS systems. Almost 
all BH--WD mergers are produced from the CE channel. About $ (10-40)\% $ of BH--NS mergers are formed through the stable mass transfer 
channel, with the merger rate densities $ \mathcal{R}_{\rm BHNS} \sim 4.9 \rm\, Gpc^{-3}yr^{-1}$,
$ 4.2  \rm\, Gpc^{-3}yr^{-1}  $ and $ 9.9  \rm\, Gpc^{-3}yr^{-1}  $ in the rapid, delayed and stochastic model, respectively. 
About $ (30-70)\% $ of the BH$ - $BH mergers form from
the stable mass transfer channel,
with $ \mathcal{R}_{\rm BHBH} \sim 29.7 \rm\, Gpc^{-3}yr^{-1}$, $ 36.6  \rm\, Gpc^{-3}yr^{-1}  $ and 
$ 22.6  \rm\, Gpc^{-3}yr^{-1}  $ in the rapid, delayed and stochastic model, respectively.
The contribution of the stable mass transfer channel stems from our revised CE criteria of mass transfer stability for the BH binaries with nondegenerate donors which allow large parameter spaces for stable mass transfer (see Section 3). 
In Figure 11, we schematically show the formation history of a BH$ - $BH merger
containing a mass-gap BH through the evolutionary channel without any CE phase. The binary evolution starts from
a primordial system consisting of a $ 30M_{\odot} $ primary star and a $ 22 M_{\odot} $ secondary star in a 10 day orbit. 
The metallicity of both stars is initially taken to be 0.001. At the time of 6.6 Myr, the primary star, which has climbed to the
supergiant branch, starts to overflow its RL. After about 0.3 Myr of stable mass transfer, the primary is stripped to
be a $\sim 10.6M_{\odot} $ Wolf-Rayet star and the secondary is rejuvenated by accretion of  $ \sim 1.8M_{\odot} $ matter. 
The mass transfer efficiency is about 0.1 during this phase, with the rotation-dependent mode. 
When the Wolf-Rayet star collapses into a BH, the binary evolves to be an eccentric 
($ e \sim0.32 $) system in a $ \sim 39 $ day orbit. At the time of 9.5 Myr, the secondary star overflows its RL and
transfers mass to the BH, causing the binary orbit to shrink rapidly. The post-mass transfer system possesses a 
$\sim 6.9M_{\odot} $ BH and a $\sim 8.5M_{\odot} $ Wolf-Rayet star in a nearly circular orbit with a period of 5.2 days.
After 0.3 Myr, a close BH$ - $BH system is formed and the second born BH has a mass of $ \sim 4.4M_{\odot} $. About 9 Gyr later,
this binary will merge to be a single BH.

\begin{figure}[hbtp]
\centering
\includegraphics[width=0.5\textwidth]{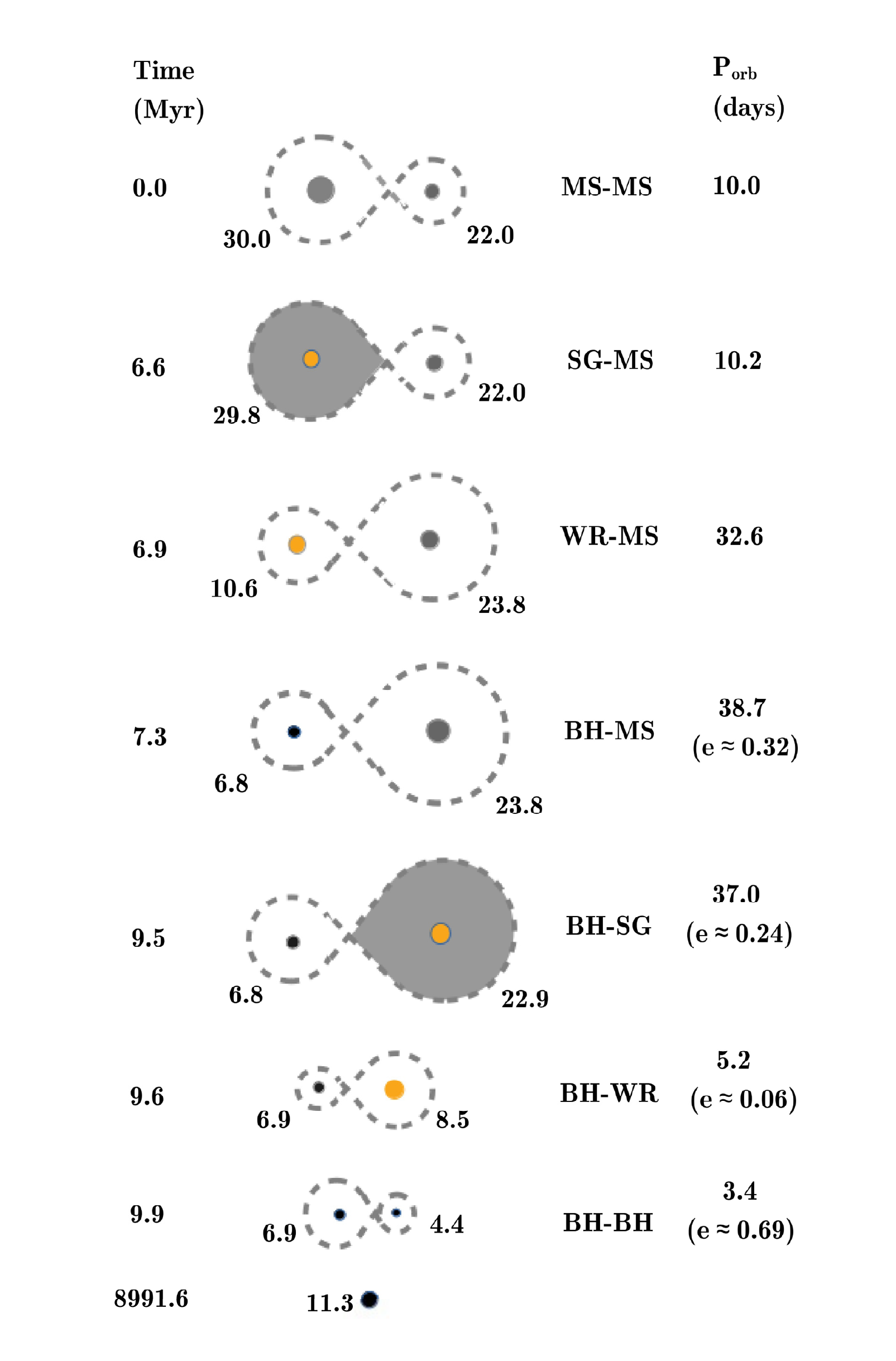}
%\linespread{0.7}
\caption{Schematic plot depicts the formation of a BH$ - $BH merger containing a mass-gap component via 
the channel without any CE phase. Acronyms for different stellar 
types used in this figure$ - $MS: main sequence; SG: supergiant; WR: Wolf Rayet. 
   \label{figure1}}
\end{figure}

\begin{figure*}[hbtp]
\centering
\includegraphics[width=0.8\textwidth]{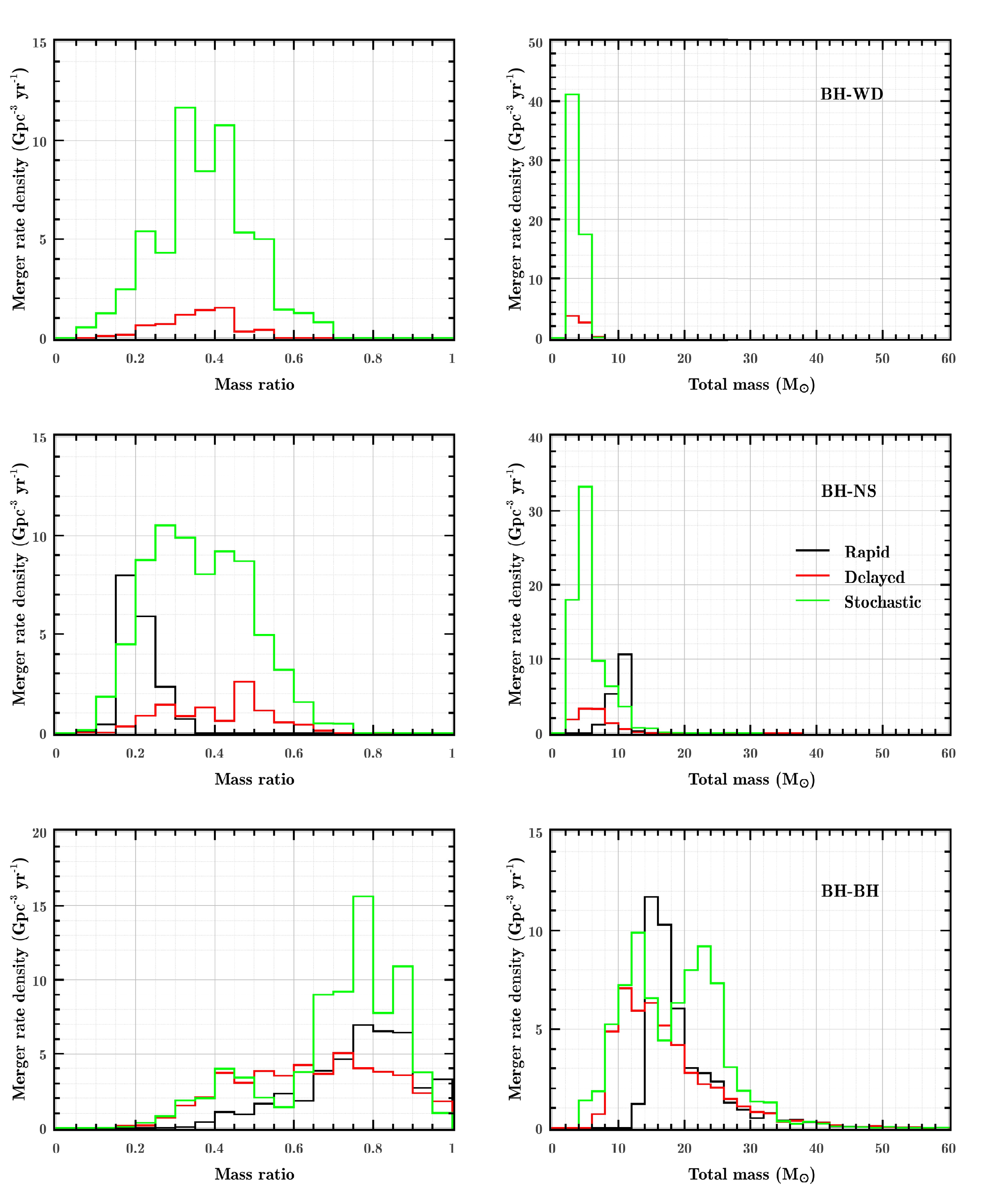}
%\linespread{0.7}
\caption{Predicted mass ratio (of the light to the heavy components) and total mass distributions of the BH$ - $CS mergers 
in the local Universe for our adopted three models. 
   \label{figure1}}
\end{figure*}

\subsection{Mass ratios and total masses of BH--CS mergers}

Figure 12 shows the merger rate density distributions of the BH$ - $CS mergers 
in the local Universe, as a function of the mass ratio (of the light to the heavy components) and the total mass. 
(1)~Since almost all BH$ - $WD mergers possess a mass-gap BH  and the WDs have mass $ \sim 1M_{\odot} $, 
both the delayed and the stochastic models predict that the mass ratios of such mergers are mainly distributed in the range of $ \sim 0.2-0.5 $. 
Meanwhile, their total masses are expected to vary in a narrow range of $ \sim 3-6 M_{\odot} $.  (2)~In the rapid model,
the BH$ - $NS mergers have the mass ratio distribution in the range of $ \sim 0.1-0.3 $ with a peak at $ \sim 0.2 $ and the 
total mass distribution in the range of $ \sim 6-12M_{\odot} $ with a peak at $ \sim 10-12M_{\odot} $. Compared with the rapid model,
the peak of the mass ratio distribution for BH$ - $NS mergers shifts to larger values of $ \sim 0.5 $ ($ \sim 0.3 $) and the peak of the total 
mass distribution to lower mass of $ \sim 4-8 M_{\odot}$ ($ \sim 4-6 M_{\odot}$) in the delayed (stochastic) model. (3)~The 
BH$ - $BH mergers in the rapid and stochastic models tend to have large mass ratios whose distribution has a broad peak at $ \sim 0.7-0.9 $
and $ \sim 0.6-0.9 $, respectively, while the delayed model has a relatively flat mass ratio distribution between $ \sim 0.4- 0.9$. 
The total masses of the BH$ - $BH mergers in the rapid model are always larger than $ \sim 12M_{\odot} $, but can extend down to
$ \sim 4-6M_{\odot} $ in the delayed and stochastic models. After the detection of GW190412, the percentage of the BH$ - $BH mergers with mass
ratios less than 0.4 has been constrained to constitute $ \gtrsim 10\% $ of the whole BH$ - $BH merger population \citep{ab20c}. We estimate
that the fractions are $ \sim 1\% $, $ \sim 9.8\% $ and $ 6.6\% $ in the rapid, delayed and stochastic models, respectively.  
We do not further discuss the distribution shape (a broken power-law function) for the component masses of BH$ - $BH mergers \citep{ab20b}, 
since one can revise the input physics (e.g. BH natal kicks and CE ejection efficiencies) to match the LIGO/Virgo data \citep[see e.g.,][]{ob21}. 

\section{Conclusions}

In this paper, we have investigated the properties of merging BH$ - $CS binary population in the Milky Way and the local Universe, 
based on binary evolution calculations with the \textit{BSE} and \textit{MESA} codes. Only the systems formed 
through isolated binary evolution are taken into account.
%implying that other formation channels may add additional contribution 
%to the synthetic population of merging BH$ - $CS binaries. Many previous works indicate that CE phase is a vital element of
%setting the formation of close BH$ - $CS systems. 
Compared with previous works, the innovations in this work mainly lie in two aspects. 
(1) We have revised the criteria for the occurrence of CE evolution in the BH binaries with nondegenerate donors, 
by a large grid of detailed binary evolution simulations with the \textit{MESA} code.
As a consequence, we obtain the potential parameter space for dynamically (un)stable mass transfer, and incorporate into
our BPS calculations. (2) We consider various mechanisms for the formation of NSs and BHs. The (non)existence of 
the mass gap between NSs and BHs is crucial to constrain the mechanism of supernova explosions. 
We adopt three models to deal with the compact remnant masses and natal kicks during supernova explosions, 
that is the rapid mechanism \citep{fb12}, the delayed mechanism \citep{fb12} and the stochastic recipe \citep{mm20}. 
The delayed and the stochastic models allow the formation of CSs within the mass gap, while the rapid model naturally leads to 
the mass gap between NSs and BHs. The fractions $ f_{\rm MG} $ of merging systems with mass-gap BHs among
the whole population for different types of BH$ - $CS systems can be used as an indicator
to examine relevant supernova mechanisms.
In the rapid model, $ f_{\rm MG} = 0$ always holds.
%We summarize our main results according to different supernova models in the following.

We identify two formation channels for close BH$ - $CS systems that appear as GW sources,
depending on whether the mass transfer in the progenitor BH binaries is stable.
We find that almost all close BH$ - $WD binaries have experienced a CE phase, during which a mass-gap BH 
was engulfed by the envelope of a $ \sim 6-10M_{\odot} $ supergiant (the WD's progenitor). 
So for merging BH$ - $WD systems, we expect that $ f_{\rm MG} \sim 1$ in both the delayed and the stochastic models. 
For merging BH$ - $NS
and BH$ - $BH systems, both the CE and stable mass transfer channels take effect. In both the delayed and stochastic models,
we obtain that $ f_{\rm MG} \sim 0.7$ for merging BH$ - $NS binaries and $ f_{\rm MG} \sim 0.3$ for merging BH$ - $BH binaries.  
It seems that our $ f_{\rm MG} $ predictions for merging BH$ - $CS binary population cannot make a clear distinction between the delayed 
and the stochastic models. The LIGO/Virgo operation will provide a rapid growing sample of BH$ - $BH merger events.
For BH$ - $BH mergers with total mass $ \lesssim 30M_{\odot} $, the delayed model anticipates a more flat mass-ratio distribution
between $ \sim 0.4-0.9 $ while the stochastic model favors large mass ratios distributed with a broad peak at $ \sim 0.6-0.9 $ (see Figure 12).

We estimate that there are totally dozens of BH$ - $CS systems detectable by LISA in the Milky Way, and 
the merger rate density of  BH$ - $CS systems varies in the range of  $ \sim 60-200 \rm\, Gpc^{-3}yr^{-1}$ in the local Universe.
For merging BH$ - $WD binaries, the delayed and stochastic models predict $ \sim 2 $ and $ \sim 38 $ systems may be 
observed by LISA in the Milky Way, respectively.  In addition, we expect that local BH$ - $WD merger rate is in the range of 
$ \sim 7-59 \rm\, Gpc^{-3}yr^{-1}$ for the delayed and stochastic models. 
Our calculations show that $ \sim 2-14 $  BH$ - $NS systems are potential GW sources in the Milky Way and the local
merger rate of  BH$ - $NS binaries is in the range of $ \sim 10-80 \rm\, Gpc^{-3}yr^{-1}$. Among all Galactic
BH$ - $BH systems, $ \sim 12-26 $ of them are expected to be the GW sources in the LISA frequency. Our calculated merger 
rate of BH$ - $BH binaries in the local Universe varies in the range of $ \sim 40-80 \rm\, Gpc^{-3}yr^{-1}$. 
At last we remind that our results are still subject to many uncertainties, including the assumption of binary fraction ($ f_{\rm b} = 1 $),
the treatments of stellar and binary evolutionary processes \citep[see e.g.,][]{ln12}, the options of Galactic and cosmological parameters, and etc. 
It is obvious that $ f_{\rm b} \sim 0.6-0.9$ from observations \citep{md17} can lead to the decrease of the LISA-detectable numbers and the
local rates for the merging BH--CS binary populations.

\acknowledgements
We thank the anonymous referee for constructive suggestions 
that helped improve this paper. This work was supported by the Natural Science Foundation 
of China (Nos.~11973026, 11773015 and 12041301), the Project U1838201 
supported by NSFC and CAS, and the National Program on Key Research and 
Development Project (Grant No. 2016YFA0400803). 

%\end{acknowledgements}

\clearpage

\end{document}